\documentclass[aps,showpacs,amsmath,amssymb,twocolumn,nofootinbib]{revtex4-1}

\usepackage{cancel}
\usepackage{cases}
\usepackage{subfigure}
\usepackage{graphicx}
\usepackage{dcolumn}
\usepackage{bm}
\usepackage{color}
\usepackage[colorlinks=true,pdfstartview=FitV,linkcolor=blue,citecolor=blue,urlcolor=blue]{hyperref}
\usepackage[mathlines]{lineno}
\usepackage[dvipsnames]{xcolor}
\usepackage{amsmath}
\usepackage{ytableau}
\usepackage{textcase}

\everymath{\displaystyle}

\begin{document}

\newcommand{\beq}{\begin{equation}}
\newcommand{\eeq}{\end{equation}}
\newcommand{\beqs}{\begin{eqnarray}}
\newcommand{\eeqs}{\end{eqnarray}}
\begin{flushright}
DESY 21-112
\end{flushright} 

\title{Probing PeV scale SUSY-breaking with
\\[0.1cm]
Satellite Galaxies and Primordial Gravitational Waves}

\author{Gongjun Choi$^{1}$}
\thanks{{\color{blue}gongjun.choi@hotmail.com}}

\author{Ryusuke  Jinno$^{2}$}
\thanks{{\color{blue}ryusuke.jinno@desy.de}}

\author{Tsutomu T. Yanagida$^{1,3}$}
\thanks{{\color{blue}tsutomu.tyanagida@sjtu.edu.cn}}

\affiliation{$^{1}$ Tsung-Dao Lee 
Institute, Shanghai Jiao Tong University, Shanghai 200240, China}

\affiliation{$^{2}$ Deutsches Elektronen-Synchrotron DESY, Notkestr. 85, 22607 Hamburg, Germany}

\affiliation{$^{3}$ Kavli IPMU (WPI), UTIAS, The University of Tokyo, 5-1-5 Kashiwanoha, Kashiwa, Chiba 277-8583, Japan}
\date{\today}

\begin{abstract}
We study an inevitable cosmological consequence in PeV scale SUSY-breaking scenarios. We focus on the SUSY-breaking scale corresponding to the gravitino mass $m_{3/2}=100{\rm eV}-1{\rm keV}$. We argue that the presence of an early matter-dominated era and the resulting entropy production are requisite for the Universe with this gravitino mass. We infer the model-independent minimum amount of the entropy production $\Delta$ by requiring that the number of dwarf satellite galaxies $N_{\rm sat}$ in the Milky Way exceed the currently observed value, i.e. $N_{\rm sat}\gtrsim63$. This entropy production is inevitably imprinted on the primordial gravitational waves (pGWs) produced during the inflationary era. We study how the information on the value of $\Delta$ and the time of entropy production are encoded in the pGW spectrum $\Omega_{\rm GW}$. 
If the future GW surveys observe a suppression feature in the pGW spectrum for the frequency range $\mathcal{O}(10^{-10}){\rm Hz}\lesssim f_{\rm GW}\lesssim\mathcal{O}(10^{-5}){\rm Hz}$, it works as a smoking gun for PeV SUSY-breaking scenarios. Even if they do not, our study can be used to rule out all such scenarios.
\end{abstract}

\maketitle
\section{Introduction}  
One of the advantages of studying gravitino cosmology lies in the fact that the gravitino mass $m_{3/2}$ is directly related to a supersymmetry (SUSY)-breaking scale $\sqrt{F}$, i.e. $m_{3/2}\simeq F/(\sqrt{3}M_{P})$ where $M_{P}\simeq2.4\times10^{18}{\rm GeV}$. For that reason, understanding effects that gravitinos can potentially have on experimental data can be of great help in studying the SUSY-breaking scale. Then, is there any inevitable physical effect induced by gravitinos so that its absence implies exclusion of a certain SUSY-breaking scale?   

Inspired by this question, in this work, we give our special attention to PeV ($10^{6} {\rm GeV}$) scale SUSY-breaking scenarios where the gravitino mass is so light that gravitinos serve as the lightest SUSY particle (LSP). Particularly, we focus on the gravitino mass range $100{\rm eV}\lesssim m_{3/2}\lesssim1{\rm keV}$.\footnote{The gravitino mass $100{\rm eV}\lesssim m_{3/2}$ is consistent with
perturbative gauge mediation models~\cite{Yanagida:2012ef}.} In this case, as far as a reheating temperature is greater than a sparticle mass, it is certain that gravitinos were once produced in the thermal bath and exist today in the form of warm dark matter (WDM) with a free-streaming length amounting to $\mathcal{O}(0.1){\rm Mpc}$.

Interestingly, for the thermal sub-keV gravitino, the relic abundance $\Omega_{3/2}h^{2}$ is insensitive to the reheating temperature $T_{\rm RH}$~\cite{Moroi:1993mb}, but sensitive to only $m_{3/2}$ and the decoupling temperature $T_{3/2,\rm dec}$. This means that once the sparticle mass spectrum is fixed, there is a definite prediction for the relic abundance of the thermal sub-keV gravitino WDM. Given the null observation of any sparticle thus far, it is fair to state that the gravitino decoupling temperature is at least $\mathcal{O}(1){\rm TeV}$. Therefore, for $100{\rm eV}\lesssim m_{3/2}\lesssim1{\rm keV}$, there exists a solid lower bound on $\Omega_{3/2}h^{2}\equiv\omega_{3/2}$ for each $m_{3/2}$ as far as $T_{\rm RH}$ is larger than sparticle masses and no entropy production occurs.

On the other hand, Lyman-$\alpha$ forest observations and the redshifted 21cm signals in EDGES observation~\cite{Bowman:2018yin} give stringent lower bounds on the WDM mass in the case where the whole of DM population consists only of a WDM component. Constraints from each experiment read $m_{\rm wdm}>5.3\,{\rm keV}$~\cite{Irsic:2017ixq} and  $m_{\rm wdm}>6.1\,{\rm keV}$~\cite{Schneider:2018xba,Lopez-Honorez:2018ipk}, respectively (See \cite{Garzilli:2019qki} for more conservative lower bound, 1.9{\rm keV}). Hence, as is well-known, sub-keV gravitino WDM cannot be responsible for 100\% DM population today if the Universe went through PeV scale SUSY-breaking. However, since these constraints assume that the whole DM is made of a single species, there is still room for the sub-keV gravitino to be part of DM if it is a subcomponent. Thus, an immediate and natural question is ``what is an upper bound on $\omega_{3/2}$ for a given $m_{3/2}\in[100{\rm eV},1{\rm keV}]$ to be consistent with cosmological data at small scales?"  

In this work, we address this question by making the estimate of the expected number of dwarf satellite galaxies $N_{\rm sat}$ in the Milky Way resulting from the Universe with the mixture of cold dark matter (CDM) and gravitino WDM. Defining the fraction of DM population contributed by the gravitino WDM to be $f_{3/2}\equiv\omega_{3/2}/\omega_{\rm DM}$, we compute $N_{\rm sat}$ based on the matter power spectrum arising from $100(1-f_{3/2})\%$ CDM and $100f_{3/2}\%$ gravitino WDM. Then by requiring $N_{\rm sat}\gtrsim63$, we obtain the maximally allowed $f_{3/2,\rm max}$ for each $m_{3/2}$. Note that we demand that the rest of the main component DM is of a cold type to avoid too much suppression of the matter growth at small scales. Furthermore, we have the CDM unspecified in our work because it depends on details of a model.

In accordance with our observation that $f_{3/2,\rm max}$ is smaller than $f_{3/2}$ expected in the universe without any mechanism to dilute the gravitino relic abundance, we argue that the entropy production is requisite for $m_{3/2}\in[100{\rm eV},1{\rm keV}]$.  We derive the necessary amount of entropy production for each $m_{3/2}$, and then argue that this entropy production is inevitably imprinted on the primordial (inflationary) gravitational wave (pGW) spectrum~\cite{Grishchuk:1974ny,Starobinsky:1979ty,Turner:1993vb,Turner:1996ck,Smith:2005mm}. This means that the feature in the pGW spectrum works as a smoking gun for PeV scale SUSY-breaking scenarios. Since the entropy production must occur after gravitino decoupling, the corresponding frequency of the pGW falls in the best frequency band explored by future experiments. Even if they detect only the pGW spectrum without any suppression feature in the frequency range $\mathcal{O}(10^{-10}){\rm Hz}\lesssim f_{\rm pGW}\lesssim\mathcal{O}(10^{-5}){\rm Hz}$, our study contributes to ruling out all the PeV scale SUSY-breaking scenarios.

The outline of this paper is as follows. In Sec.~\ref{sec:sec2}, we raise questions we address in this work by making a brief review for $m_{3/2}=\mathcal{O}(100){\rm eV}$ gravitino cosmology. Then in Sec.~\ref{sec:deltaNsat}, we discuss how to obtain the maximally allowed $f_{3/2,\rm max}$ for each $m_{3/2}$ and discuss the result of the analysis. In Sec.~\ref{sec:GW}, we discuss inevitable signatures on the pGW spectrum with PeV scale SUSY-breaking. We study the deviation of the pGW spectrum from its usual, almost scale-invariant form induced by the presence of an early matter dominated (EMD) era followed by the entropy production via the decay of a heavy degree of freedom. Finally our conclusion is made in Sec.~\ref{sec:conclusion}.


\section{P\texorpdfstring{\MakeLowercase{e}V}{eV} scale SUSY Breaking Scenario}
\label{sec:sec2} 
In a local SUSY model, the SUSY-breaking scale $M_{\cancel{\rm SUSY}}=\sqrt{|F|}$ is directly connected to the gravitino mass ($m_{3/2}$)  via 
\beq
m_{3/2}=\frac{F}{\sqrt{3}M_{P}}\,.
\label{eq:m3/2}
\eeq
Particularly, for a low SUSY-breaking scale $F\simeq10^{11}-10^{12}{\rm GeV^{2}}$ ($M_{\cancel{\rm SUSY}}=\mathcal{O}(1){\rm PeV}$), the corresponding gravitino mass is found to be $m_{3/2}\in[100{\rm eV},1{\rm keV}]$.

This gravitino mass range is of particular interest in that the relic abundance of the gravitinos is given almost in a model-independent way. On the spontaneous SUSY-breaking, gravitinos  become massive by absorbing the Goldstino field. As far as sparticles are present in the MSSM thermal bath, gravitinos remain in equilibrium with the thermal bath. Afterwards, the thermal bath continues to cool down until its temperature ($T$) reaches the following gravitino decoupling temperature $T_{3/2,{\rm dec}}$~\cite{Moroi:1993mb}\footnote{{\rm Max}[A,B] means the larger one among A and B.}  
\beq
{\rm Max}\!\left[m_{\tilde{g}},10{\rm GeV}\left(\frac{g_{*\rho}(T_{3/2,{\rm dec}})}{230}\right)^{\!\frac{1}{2}}\left(\frac{m_{3/2}}{1{\rm keV}}\right)^{\!2}\left(\frac{1{\rm TeV}}{m_{\tilde{g}}}\right)^{\!2}\right],
\label{eq:Td}
\eeq
where $g_{*\rho}(T)$ is the effective number of relativistic degrees of freedom for the energy density in the MSSM thermal bath at temperature $T$, and $m_{\tilde{g}}$ is the gluino mass.

For $F\simeq10^{11}-10^{12}{\rm GeV^{2}}$, the low scale gauge-mediated SUSY-breaking (GMSB) can explain soft masses of sfermions and gauginos. The gluino mass is dominantly generated by the loop correction contributed from colored messengers, which reads
\beq
m_{\tilde{g}}\simeq N_{\rm mess}\frac{g_{c}^{2}}{(4\pi)^{2}}\frac{yF}{M_{\rm mess}}\,,
\label{eq:gauginomass}
\eeq
where $N_{\rm mess}$ is the number of messengers, $g_{c}$ is the gauge coupling of the MSSM $SU(3)_{c}$ color gauge group, $y$ is the coupling constant for the interaction between messenger fields and SUSY-breaking field, and $M_{\rm mess}$ is the mass of the messenger. 

For a perturbative gauge mediation model, in order for the SUSY-breaking vacuum to be stable to date, the condition $M_{\rm mess}^{2}>\!\!>yF$ needs to be satisfied~\cite{Hisano:2008sy}. This implies that $m_{\tilde{g}}$ can be at most $\mathcal{O}(10){\rm TeV}$ for $m_{3/2}\in[100{\rm eV},1{\rm keV}]$. Thus, we see that $T_{3/2,{\rm dec}}\simeq m_{\tilde{g}}$ holds true in Eq.~(\ref{eq:Td}). For a given value of $m_{3/2}$, $T_{3/2,{\rm dec}}$ thus obtained leads to the following estimate of the relic abundance of gravitinos
\beqs
\omega_{3/2}\equiv\Omega_{3/2}h^{2}&=&\left(\frac{T_{3/2,0}}{T_{\nu,0}}\right)^{\!3}\left(\frac{m_{3/2}}{94{\rm eV}}\right)\cr\cr
&=&\left(\frac{10.75}{g_{*s}(T_{3/2,{\rm dec}})}\right)\left(\frac{m_{3/2}}{94{\rm eV}}\right)\,,
\label{eq:omega32}
\eeqs
where $\Omega_{3/2}$ is the ratio of the present gravitino energy density to the critical energy, $h$ parametrizes the Hubble expansion rate via $H_{0}=100h{\rm km/Mpc/sec}$, $g_{*s}(T)$ is the effective number of relativistic degrees of freedom for the entropy density in the MSSM thermal bath at temperature $T$, and $T_{3/2,0}$ and $T_{\nu,0}$ are the temperature of gravitinos and neutrinos at present, respectively.

Since $g_{*s}(T_{3/2,{\rm dec}})$ depends on a model-dependent mass spectrum, we cannot obtain a one-to-one correspondence between $\omega_{3/2}$ and $m_{3/2}$ from Eq.~(\ref{eq:omega32}). However, because $g_{*s}$ in the MSSM is at most $g_{*s}\simeq230$,\footnote{Assuming GMSB scenario, $g_{*s}$ can be slightly greater than $g_{*s}\simeq230$ due to additional contribution made by messengers.} it is fair to state that for each $m_{3/2}$ the minimum inevitable value of $\Omega_{3/2}h^{2}$ is given by
\beq
\omega_{3/2}\gtrsim\omega_{3/2,{\rm min}}=\left(\frac{10.75}{230}\right)\left(\frac{m_{3/2}}{94{\rm eV}}\right)\,.
\label{eq:omega32min}
\eeq
Given $\omega_{3/2,{\rm min}}$ in Eq.~(\ref{eq:omega32min}), we notice that for the Universe with $m_{3/2}\in[100{\rm eV},1{\rm keV}]$, as the lightest SUSY particle (LSP), the gravitinos must exist today in the form of DM. Depending on its mass, the gravitino's relic abundance can explain a fraction of DM today or even exceed the DM relic abundance in the absence of any dilution mechanism. Making the estimate of the free-streaming length ($\lambda_{\rm FS}$) of gravitinos via~\cite{Viel:2005qj},
\beq
\lambda_{\rm FS}\sim\frac{2\pi}{5}\left(\frac{m_{3/2}}{1{\rm keV}}\right)^{\!-1}\left(\frac{10.75}{g_{*s}(T_{3/2,{\rm dec}})}\right)^{\!\frac{1}{3}}{\rm Mpc}\,,
\label{eq:FS}
\eeq
it is realized that sub-keV gravitinos with $g_{*s}\simeq230$ serve as WDM owing to $\lambda_{\rm FS}\sim\mathcal{O}(0.1)-\mathcal{O}(1){\rm Mpc}$.

Therefore, if our Universe went through the spontaneous SUSY-breaking at PeV scale, it becomes inevitable today for the DM population to be composed of a CDM candidate and gravitino WDM (mixed DM scenario) or fully of gravitino WDM with a certain mechanism to dilute the relic abundance. If so, one important question to be addressed is whether the presence of gravitino WDM is consistent with various cosmological and astrophysical observables or not.

Regarding this question, for the later possibility, already the mass constraints on WDM from Lyman-$\alpha$ forest observation $m_{\rm wdm}^{\rm thermal}>5.3\,{\rm keV}$~\cite{Irsic:2017ixq} and from the redshifted 21cm signals in EDGES observations $m_{\rm wdm}^{\rm thermal}>6.1\,{\rm keV}$~\cite{Schneider:2018xba,Lopez-Honorez:2018ipk} require the dilution of the energy density. This implies that for $m_{3/2}\in[100{\rm eV},1{\rm keV}]$, gravitinos can reside in the present Universe only as a fraction of the DM population. 

Along this line of reasoning, we ask two key questions that have a direct connection to cosmological consequences of PeV SUSY-breaking scenarios and provide us with a powerful cosmological probe of PeV SUSY-breaking scenarios. These questions are
\begin{enumerate}
    \item For each $m_{3/2}$, what is the minimum amount of the entropy production to make the scenario consistent with observations?
    \item How to probe and confirm such an entropy production?
\end{enumerate}

In Sec.~\ref{sec:deltaNsat} and Sec.~\ref{sec:GW}, we answer these questions by invoking the estimate of the number of satellite galaxies in the Milky Way (MW) and pGWs. In Sec.~\ref{sec:deltaNsat}, we quantify the necessary amount of the entropy production by $\Delta\equiv S/\overline{S}$ where $\overline{S}$ and $S$ are the total entropy before and after the decay of a heavy particle $X$, respectively, which we assume to take place prior to Big Bang Nucleosynthesis (BBN) era.\footnote{In order not to spoil successful BBN, we assume that the entropy production takes place before the temperature drops down to $10{\rm MeV}$. As explained later, we consider entropy production from the decay of a heavy particle $X$ whose decay rate determines the decay time (temperature) via $\Gamma_{X}\simeq H$.
Possible candidates of $X$ include particles in the messenger sector in GMSB scenarios~\cite{Fujii:2002fv}, particles in the SUSY-breaking sector~\cite{Ibe:2010ym}, or the lightest right-handed sneutrino~\cite{Choi:2021uhy}.} To this end, we use the observed number of the satellite galaxies $N_{\rm sat}$ in the MW to obtain the maximal allowed fraction of gravitino for each $m_{3/2}$. This eventually provides us with the minimally required dilution factor $\Delta_{\rm min}$. In Sec.~\ref{sec:GW}, we study the imprint of the EMD era and the entropy production on the pGW spectrum. Regarding testability, we discuss how future GW detection experiments can be employed to either confirm or rule-out PeV-scale SUSY breaking scenarios.


\section{$\Delta_{\rm min}$ from $N_{\rm sat}$}
\label{sec:deltaNsat} 
The mixed DM model (MDM), in which the present DM population consists of both CDM and WDM, is parametrized by two quantities: the WDM mass and the fraction of the DM relic abundance today attributable to WDM. For our case with gravitinos being WDM, these are denoted by $(m_{3/2},f_{3/2})$. For a given set of $(m_{3/2},f_{3/2})$, the matter power spectrum in MDM scenario ($P_{\rm MDM}(k)$) can be parametrized as
\beq
P_{\rm MDM}(k)=\mathcal{T}(k,m_{3/2},f_{3/2})^{2}P_{\rm CDM}(k)\,,
\label{eq:Pmdm}
\eeq
where $\mathcal{T}(k,m_{3/2},f_{3/2})$ is a transfer function.

\begin{figure}[t]
\centering
\hspace*{-5mm}
\includegraphics[width=0.49\textwidth]{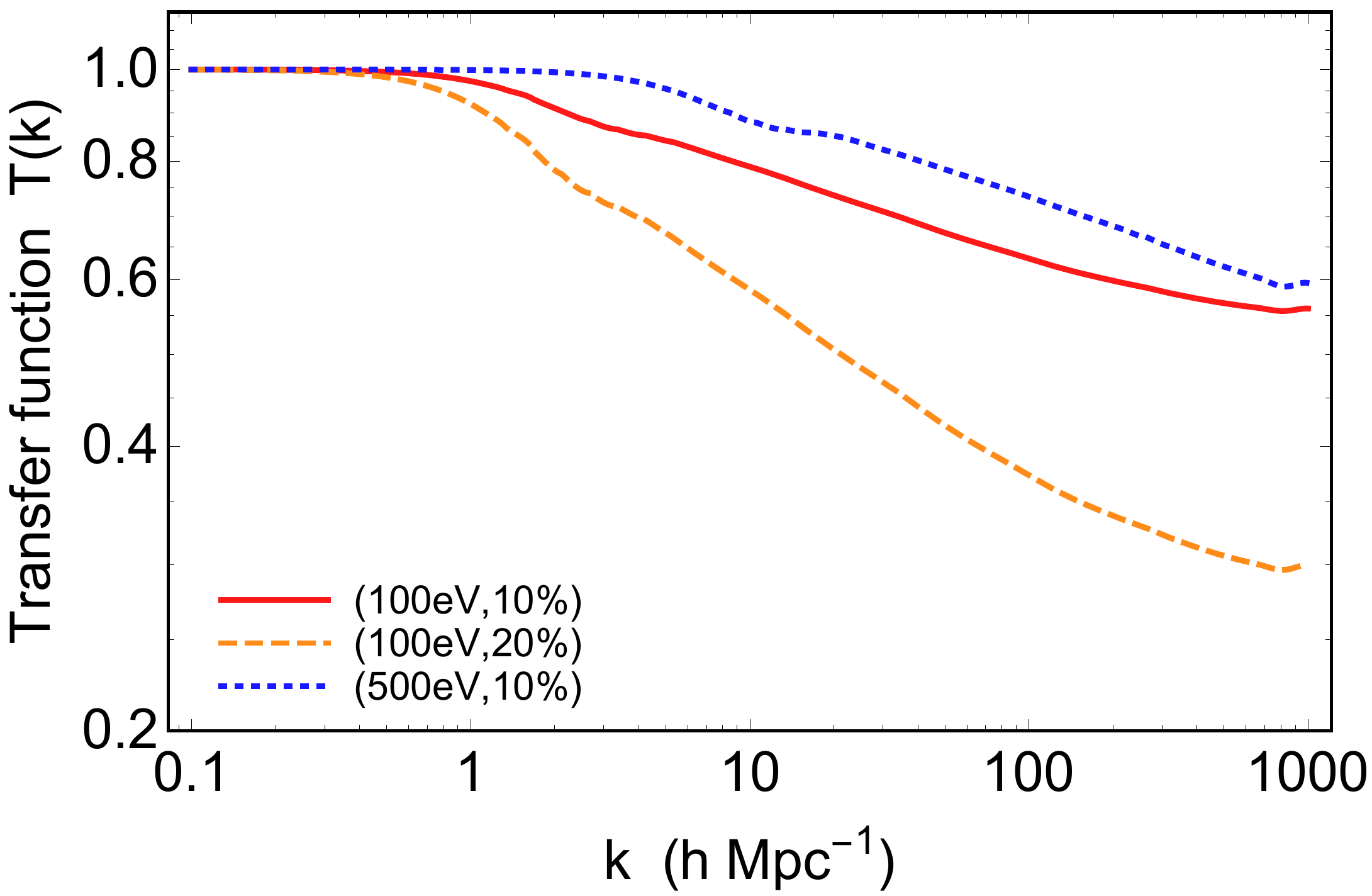}
\caption{Transfer function defined in Eq.~(\ref{eq:Pmdm}) for different values of $(m_{3/2},f_{3/2})$.}
\vspace*{-1.5mm}
\label{fig:1}
\end{figure}

Relating the matter power spectrum in the CDM case ($P_{\rm CDM}(k)$) to the one in the MDM case ($P_{\rm MDM}(k)$), the transfer function contains information for a mass and WDM fraction in a MDM scenario. Suppressing $P_{\rm CDM}(k)$ at small scales, the transfer function is characterized by two quantities   $(m_{3/2},f_{3/2})$. Its structure is determined by (1) a comoving wavenumber $k_{\rm sup}$ beyond which the transfer function starts to deviate from the unity and (2) the depth of suppression $\mathcal{T}_{\rm plateau}$ at small scales (for $k>k_{\rm sup}$)~\cite{Boyarsky:2008xj}. $k_{\rm sup}$ is parametrized by the temperature ratio in Eq.~(\ref{eq:omega32}) and $m_{3/2}$ while $\mathcal{T}_{\rm plateau}$ is done by $f_{3/2}$. This can be better understood via Fig.~\ref{fig:1}. By comparing the red solid line and the yellow dashed line, one can see that a larger $f_{3/2}$ induces a greater suppression. Moreover, comparison between the red solid line and the blue dotted line shows that $k_{\rm sup}$ becomes smaller for the smaller $m_{3/2}$. 

We compute $P_{\rm MDM}(k)$ for a given set of $(m_{3/2},f_{3/2})$ by using the Boltzmann solver \texttt{CLASS}~\cite{Blas:2011rf}. For our purpose, we rely on variation of the parameters in \texttt{ncdm} sector of \texttt{CLASS}. For a given $m_{3/2}$, by varying $\Omega_{\rm CDM}h^{2}$ and $\Omega_{\texttt{ncdm}}h^{2}$, we set $f_{3/2}$ which we aim for. Also for the case with $g_{*s}\neq230$, we set a present gravitino temperature to be the one deduced from $f_{3/2}$ and Eq.~(\ref{eq:omega32}) via the parameter $T_{\texttt{ncdm}}$ in \texttt{CLASS}.

As one of the ways to constrain $(m_{3/2},f_{3/2})$, we attend to the estimate of the number of dwarf satellite galaxies in the Milky Way. The approach we adopt in this paper is based on the one given in Ref.~\cite{Polisensky:2010rw} (see also Refs.~\cite{Giocoli:2007gf,Maccio:2009isa,Horiuchi:2013noa,Kennedy:2013uta,Schneider:2014rda,Schneider:2016uqi,Gariazzo:2017pzb,Diamanti:2017xfo,DEramo:2020gpr}): Fifteen satellite galaxies were observed by SDSS (Sloan Digital Sky Survey) with the sky coverage $f_{\rm sky}\simeq0.28$. When this limited sky coverage and eleven classically known satellites are taken into account together, $N_{\rm sat}\simeq63$ is inferred as the total number of satellites of the Milky Way. Considering the possibility that more satellites will be found in the future surveys, we take $N_{\rm sat}\simeq63$ as a lower bound for the number of the satellite galaxies that any DM model should satisfy. This set-up gives us a upper bound on $f_{3/2}$ for each $m_{3/2}$ below which the presence of gravitinos is consistent with the number of satellite galaxies of the Milky Way.

Given $P_{\rm MDM}(k)$, one can make the estimate of the expected number of dwarf satellite galaxies residing in a host halo.
Our estimate closely follows Refs.~\cite{Schneider:2014rda,Schneider:2016uqi}, which are based on the extended Press-Schechter approach~\cite{Press:1973iz,Bond:1990iw} with the conditional mass function~\cite{Lacey:1993iv}.
Adopting a sharp-$k$ filter for the window function, it is calculated by~\cite{Schneider:2014rda,Schneider:2016uqi}
\beq
N_{\rm sat}=\int_{M_{\rm min}}^{M_{\rm h}} dM_{\rm s}\frac{1}{C_{n}}\frac{1}{6\pi^{2}R_{\rm s}^{3}}\left(\frac{M_{h}}{M_{\rm s}^{2}}\right)\frac{P_{\rm MDM}(1/R_{\rm s})}{\sqrt{2\pi(S_{\rm s}-S_{\rm h})}}\,,
\label{eq:Nsat}
\eeq

where $M_{i}$, $R_{i}$, and $S_{i}$ are the mass, the filter scale, and the variance of the satellite galaxy (for $i=s$) or of the host halo (for $i=h$), respectively.
While the relation between the mass $M_{i}$ and the filter scale $R_{i}$ is in principle unconstrained in the sharp-$k$ modeling, we follow Ref.~\cite{Schneider:2014rda} and adopt $M_{i}=(4\pi/3)\times(c R_{i})^{3}\times\Omega_{m}\times\rho_{\rm cr,0}$ with $c=2.5$, which matches with observations best.
Here $\rho_{\rm cr,0}$ is the critical density. The overall normalization is taken to be $C_{n}=45$ to reproduce the N-body simulation result~\cite{Lovell:2013ola}.
In addition, we take $M_{\rm min}=10^{8}h^{-1}M_{\odot}$ as the minimum mass of the dwarf satellite galaxies~\cite{Brooks:2012vi} and $M_{h}=1.77\times10^{12}h^{-1}M_{\odot}$ as the Milky Way mass based on Ref.~\cite{Guo:2009fn}. The variance $S_{i}$ is the function of $R_{j}$ and given by
\beq
S_{i}=\frac{1}{2\pi^{2}}\int_{0}^{1/R_{i}}dk \; k^{2}P_{\rm MDM}(k)\,.
\label{eq:Si}
\eeq

\begin{figure*}[htp]
  \centering
  \hspace*{-5mm}
  \subfigure{\includegraphics[scale=0.5]{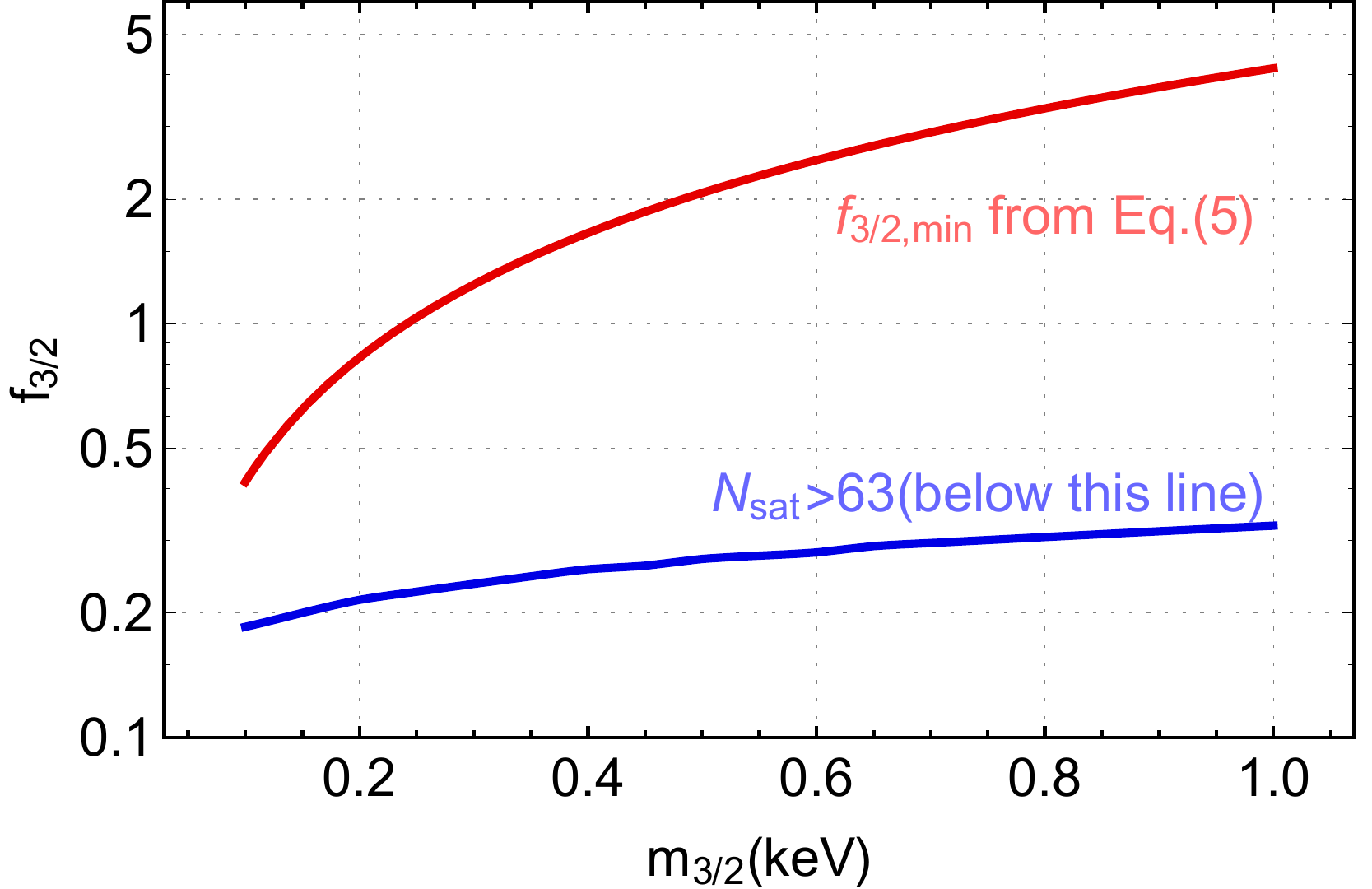}}\qquad
  \subfigure{\includegraphics[scale=0.5]{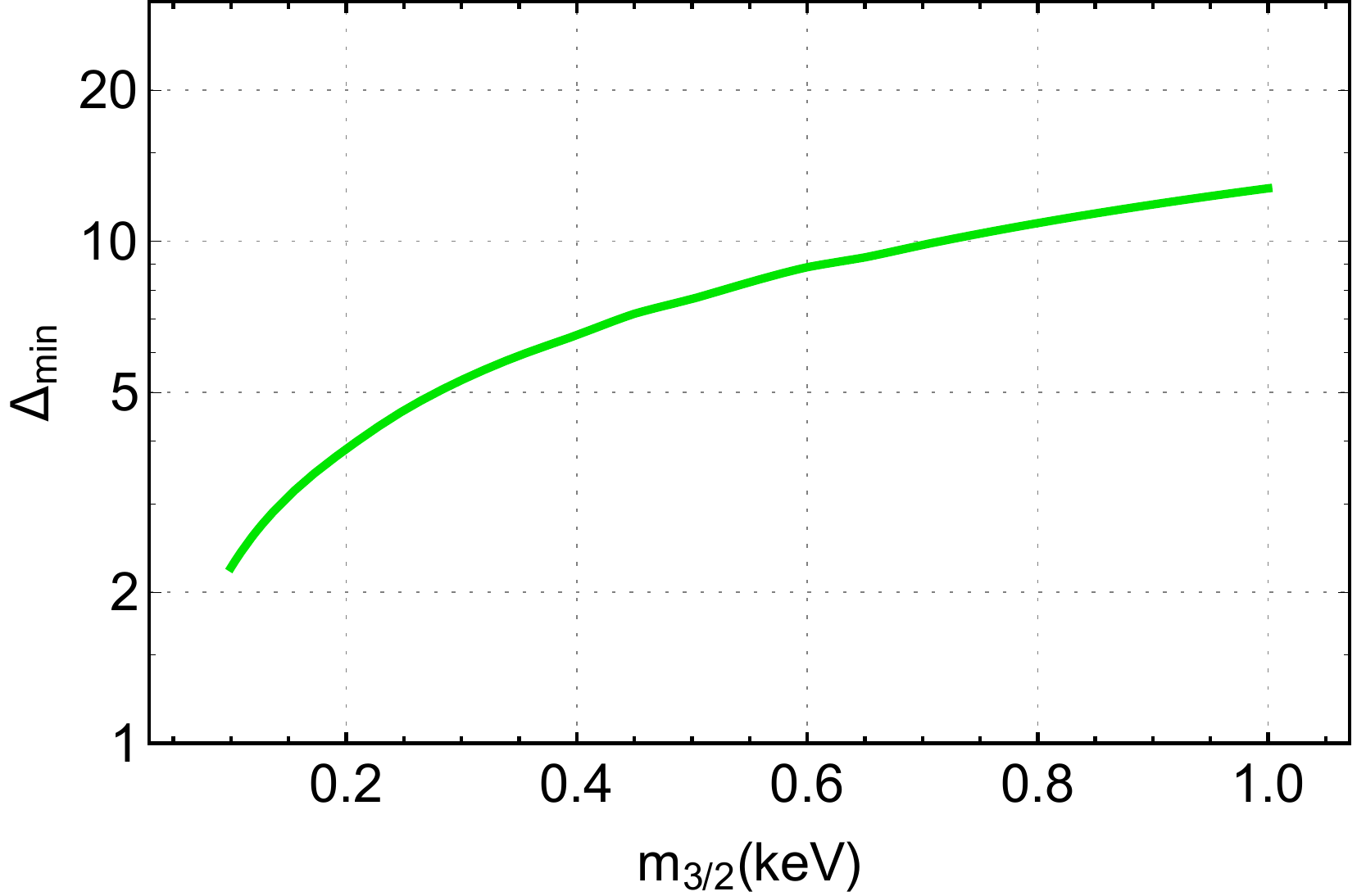}}
  \caption{{\bf Left panel}: For each gravitino mass ($m_{3/2}$) the red line shows the minimum unavoidable fraction ($f_{3/2,{\rm min}}$) of DM contributed by gravitino WDM based on Eq.~(\ref{eq:omega32min}). The region below the blue line shows $f_{3/2}$ satisfying the requirement $N_{\rm sat}>63$. {\bf Right panel}: The green line shows the minimum required amount of entropy production $\Delta\equiv S/\bar{S}$ obtained by dividing the values on the red line by those on the blue line in the left panel for each mass. $S$ is the total entropy after a heavy particle decay (entropy production) while $\bar{S}$ is the existing entropy before the entropy production took place.}
  \vspace*{-1.5mm}
\label{fig:2}
\end{figure*}

In the left panel of Fig.~\ref{fig:2}, using Eq.~(\ref{eq:omega32min}), we show for each $m_{3/2}$ (i) the minimum fraction of DM contributed by the gravitino WDM which is unavoidable in PeV SUSY-breaking scenarios in the absence of the energy dilution (red line), and (ii) the fraction of gravitino satisfying $N_{\rm sat}=63$ (blue line). Below the blue line, the number of satellite galaxies is larger than 63. As one can see from the gap between the two lines, even the minimum predicted amount of gravitino exceeds the observationally allowed one. This implies that our Milky Way should be left with too few satellites if there is no any history of dilution of the gravitino energy density. We take this point, therefore, as a strong hint for the presence of an era when the entropy production is made via, for example, a heavy particle decay.

Given Fig.~\ref{fig:1} and Eq.~(\ref{eq:Nsat}), now one can grasp in a qualitative manner how the parameters $(m_{3/2,f_{3/2}})$ have an effect on $N_{\rm sat}$. As seen in the integral range in Eq.~(\ref{eq:Nsat}), we consider the mass range $10^{8}h^{-1}M_{\odot}-10^{12}h^{-1}M_{\odot}$ as the mass range of the satellite galaxies. This mass range corresponds to the range of the filter scale $0.03h^{-1}{\rm Mpc}<R_{s}<0.66h^{-1}{\rm Mpc}$. For $m_{3/2}\in[100{\rm eV},1{\rm keV}]$, as can be seen in Fig.~\ref{fig:1}, the suppression in $P_{\rm MDM}$ as compared to $P_{\rm CDM}$ starts to occur for $\mathcal{O}(0.1)<k_{\rm sup}/(h {\rm Mpc}^{-1})<\mathcal{O}(10)$. With $k=R_{s}^{-1}$, now these imply that suppression in $P_{\rm MDM}$ induced by the presence of sub-keV gravitinos reduces $N_{\rm sat}$ by making the integrand in Eq.~(\ref{eq:Nsat}) smaller than CDM case. Because the smaller $m_{3/2}$ and the large $f_{3/2}$ give rise to the larger suppression in $P_{\rm MDM}$, we expect that the criterion $N_{\rm sat}>63$ yields more stringent constraint on $f_{3/2}$ for a smaller $m_{3/2}$.

In the right panel of Fig.~\ref{fig:2}, we show the minimum necessary amount of entropy production for the scenario to be consistent with the observed number of the satellite galaxies of the Milky Way. For each $m_{3/2}$, $\Delta_{\rm min}=S/\bar{S}$ is obtained by dividing $f_{3/2,{\rm min}}$ (red line) by $f_{3/2}$ associated with $N_{\rm sat}=63$ (blue line) in the left panel of Fig.~\ref{fig:2}. Note that when the precise $g_{*s}(T_{3/2,{\rm dec}})$ is taken into account, a larger $\Delta$ would be required. For $g_{*s}(T_{3/2,{\rm dec}})<230$, the red line moves upward while the blue line does downward, causing the green line to go up.

Before leaving Fig.~\ref{fig:2}, one may wonder whether the Lyman-$\alpha$ forest observations can offer a stronger constraint on $f_{3/2}$ than shown in the blue line in the left panel of Fig.~\ref{fig:2}. Indeed in Ref.~\cite{Murgia:2018now}, it is pointed out that the MDM model is very difficult to accommodate Lyman-$\alpha$ data unless $f_{3/2}$ is tiny. Provided $f_{3/2}$ is more severely constrained by Lyman-$\alpha$ forest data,\footnote{For instance, for a WDM with $m_{\rm wdm}\gtrsim700{\rm eV}$, it was argued in Ref.~\cite{Baur:2017stq} that $f_{\rm wdm}\sim0.15 (\sim0.1)$ can be consistent with the Lyman-$\alpha$ data (and higher-resolution data). } apparently more entropy production is required than the green line in the right panel of Fig.~\ref{fig:2} for each $m_{3/2}$. In any case, the green line remains as the most conservative lower bound for the necessary amount of $\Delta$ to be consistent with a variety of cosmological observations.

Given this inevitable requirement any PeV SUSY-breaking scenario confronts, the next important question is how one can confirm the presence of the EMD era and the sudden increase in the radiation energy density. In the next section, we address this question by studying the pGW spectrum.
We will see that these non-standard histories leave their trace on the tensor modes produced from the quantum fluctuation during inflation and re-entering the horizon before or during the entropy production.

We envision a scenario where a heavy particle $X$ is present in the first radiation-dominated (RD1) era. $X$ is assumed to be out-of-equilibrium from the thermal bath and thus its energy density scales as $\rho_{X}\propto a^{-3}$. When $\rho_{X}$ becomes comparable to the energy density of the existing thermal bath, an EMD era starts and continues until $X$ decays to produce the additional entropy. After the decay, the second radiation-dominated (RD2) era gets started and continues until the matter-radiation equality at $z_{\rm eq}\simeq3300$ is reached.


\section{Primordial Gravitational Waves}
\label{sec:GW} 
Gravitational waves $h_{ij}(t,\bold{x})$ in the Friedmann-Robertson-Walker (FRW) background can be written as
\beq
ds^{2}=a(\tau)^{2}[-d\tau^{2}+(\delta_{ij}+h_{ij})dx^{i}dx^{j}]\quad(i,j=1,2,3)\,,
\label{eq:metric}
\eeq
where $\tau$ is the conformal time defined via $dt=ad\tau$, and $h_{ij}$ satisfies the traceless and transverse conditions $h_{ii}=\partial^{i}h_{ij}=0$. In the Fourier space, $h_{ij}(t,\bold{x})$ is written as
\beqs
h_{ij}(t,\bold{x})&=&\sum_{\lambda=+,\times}\int\frac{d^{3}k}{(2\pi)^{3/2}}h_{ij}(t,\bold{k})e^{i\bold{k}\cdot\bold{x}}\,,\label{eq:hij}\\
h_{ij}(t,\bold{k})&=&h_{\bold{k}}^{\lambda}(t)\epsilon_{ij}^{\lambda}(\hat{\bold{k}})\,,
\eeqs
where $\lambda = +, \times$ are the GW polarizations, $\hat{\bold{k}}$ is the unit vector along the three momentum $\bold{k}$, and the polarization tensors $\epsilon_{ij}^{\lambda}$ are traceless and transverse as with $h_{ij}(t,\bold{x})$. The polarization tensors are normalized via $\epsilon_{ij}^{\lambda}(\epsilon_{ij}^{\lambda'})^{*}=2\delta^{\lambda\lambda'}$. The Fourier components $h_{\bold{k}}^{+}$ and $h_{\bold{k}}^{\times}$ commonly obey the following time evolution equation in the absence of an anisotropic stress
\beq
\ddot{h}_{\bold{k}}^{\lambda}+3H\dot{h}_{\bold{k}}^{\lambda}+\frac{k^{2}}{a^{2}}h_{\bold{k}}^{\lambda}=0\,,
\label{eq:hTEeqn}
\eeq
where the dot denotes the time derivative. 

After the relevant modes get sufficiently sub-horizon, the energy density of pGWs is given as
\beq
\rho_{\rm GW}=\frac{1}{32\pi G}\frac{\langle(\partial h_{ij}/\partial\tau)^{2}\rangle}{a^{2}}\,,
\label{eq:rhoGW}
\eeq
where $G$ $(\equiv (8\pi M_P^2)^{-1})$ is the Newtonian constant and $\langle...\rangle$ denotes the oscillation average. In this paper we assume the homogeneity and isotropy of the pGWs. As long as we average over a sufficiently large number of oscillations, we may add the spatial average to the definition of $\langle...\rangle$. Using the Fourier transform of $h_{ij}$ in Eq.~(\ref{eq:hij}), $|\partial h^{\lambda}_{\bold{k}}(\tau)/\partial\tau|^{2}\simeq k^{2}|h_{\bold{k}}(\tau)|^{2}$ for the sub-horizon $k$-modes and translating the spatial average into the ensemble average, one obtains the pGW energy density per logarithmic wavenumber 
\beqs
\Omega_{\rm GW}(k,\tau)&\equiv&\frac{1}{\rho_{\rm tot}}\frac{d\rho_{\rm GW}}{d\ln k}
=\frac{1}{12}\left(\frac{k}{aH}\right)^{\!2}\mathcal{P}_{T}(k,\tau)\,,
\label{eq:OmegaGW}
\eeqs
where we define $k=\sqrt{\bold{k}\cdot\bold{k}}$. The label $\tau$ is understood as the time coordinate after taking oscillation average around it. The tensor power spectrum $\mathcal{P}_{T}(k,\tau)$ is defined via the ensemble average as
\beqs
\langle h_{ij}(\tau,\bold{k})h_{ij}(\tau,\bold{k}')\rangle&=&\delta^3(\bold{k}-\bold{k}')\mathcal{P}_{T}(k,\tau)\,,
\eeqs
which can be decomposed into the primordial part and the transfer function encoding the time evolution
\beqs
\mathcal{P}_{T}(k,\tau)&=&\mathcal{P}_{T}^{\rm prim}(k)\mathcal{T}_{T}^{2}(k,\tau)\,.
\label{eq:twopointhij}
\eeqs
The transfer function is defined through $h_{\bold{k}}^{\lambda}(\tau)=h_{\bold{k}}^{\lambda,{\rm prim}}\mathcal{T}_{T}(k,\tau)$. For the primordial part, $\mathcal{P}_{T}^{\rm prim}(k)$ is written in terms of the CMB pivot scale $k_{\rm CMB}=0.002{\rm Mpc}^{-1}$ as
\beq
\mathcal{P}_{T}^{\rm prim}(k)=r\mathcal{P}_{\mathcal{R}}^{\rm prim}(k_{\rm CMB})\left(\frac{k}{k_{\rm CMB}}\right)^{\!n_{T}}\,,
\label{eq:Tpowerspectrum}
\eeq
where $\mathcal{P}_{\mathcal{R}}^{\rm prim}(k_{\rm CMB})\simeq2.1\times10^{-9}$ is the amplitude of the scalar perturbation spectrum, $r\equiv\mathcal{P}_{T}^{\rm prim}(k_{\rm CMB})/\mathcal{P}_{\mathcal{R}}^{\rm prim}(k_{\rm CMB})$ is the tensor-to-scalar ratio and $n_{T}$ is the tensor spectral index. Note that the most recent bound on the tensor-to-scalar ratio is $r\lesssim0.056$~\cite{Akrami:2018odb} which gives $\mathcal{P}_{T}^{\rm prim}(k_{\rm CMB})\lesssim1.176\times10^{-10}$.

While $\mathcal{P}_{T}^{\rm prim}(k)$ is determined by the initial conditions set during the inflationary era as in Eq.~(\ref{eq:Tpowerspectrum}), $\mathcal{T}_{T}^{2}(k,\tau)$ reflects physical processes experienced by the $k$-mode after horizon re-entry. Even within the Standard Model, the pGW spectrum carries rich information on the particle content in/out of the thermal bath~\cite{Schwarz:1997gv,Watanabe:2006qe,Kuroyanagi:2008ye,Saikawa:2018rcs}. In BSM scenarios we expect much richer structures~\cite{Seto:2003kc,Boyle:2005se,Jinno:2011sw,Jinno:2012xb,Jinno:2013xqa,Caldwell:2018giq,Ringwald:2020ist,Ringwald:2020vei}. Similarly in the present scenario, the cosmological effects of the PeV scale SUSY-breaking is encoded in $\mathcal{T}_{T}^{2}(k,\tau)$ through the entropy production~\cite{Nakayama:2008ip,Kuroyanagi:2011fy,Jinno:2014qka,Kuroyanagi:2014nba,DEramo:2019tit}.
The decay of the pGW amplitude inversely proportional to the scale factor after horizon re-entry implies $\mathcal{T}_{T}^{2}(k,\tau)\propto a^{-2}$. For the modes re-entering the horizon during deep in either RD1 or RD2 era, we can write the transfer function as~\cite{Boyle:2005se}\footnote{
Depending on the equation of state at the time of horizon re-entry, an extra factor appears in Eq.~(\ref{eq:transferF}) due to the non-zero ``thickness" of the re-entry~\cite{Boyle:2005se}. For the modes re-entering during a RD era, this extra factor takes unity and Eq.~(\ref{eq:transferF}) holds true.
}
\beqs
\mathcal{T}_{T}^{2}(k,\tau)&=&\frac{1}{2}\left(\frac{a_k}{a}\right)^{\!2}\,,
\label{eq:transferF}
\eeqs
where the subscript $k$ denotes the time of re-entry $k=a_kH_k$, and the factor $1/2$ arises from the oscillation average deep inside the horizon. Thus we obtain the following pGW spectrum expression that holds for these modes
\beq
\Omega_{\rm GW}(k,\tau)=\frac{1}{24}\left(\frac{a_k}{a}\right)^{\!4}\left(\frac{H_k}{H}\right)^{\!2}\mathcal{P}_{T}^{\rm prim}(k)\,.
\eeq
We further use the Friedmann equation $3M_P^2H_k^{2}=\rho_{\rm rad}=g_{*\rho,k}(\pi^2/30)T_k^4$ at the time of horizon re-entry together with the entropy relation $g_{*s,k}a_k^3T_k^3=\Delta^{-1}\times g_{*s,0}a_0^3T_0^3$ (for the modes re-entering during RD1) or $g_{*s,0}a_0^3T_0^3$ (RD2) to obtain the present pGW spectrum
\beqs
\Omega_{\rm GW,0}(k)&=&\frac{\Omega_{\rm rad,0}}{24}\left(\frac{g_{*\rho,k}}{g_{*\rho,0}}\right)\left(\frac{g_{*s,0}}{g_{*s,k}}\right)^{\!\frac{4}{3}}\mathcal{P}_{T}^{\rm prim}(k)\notag\\&&\times\left\{\begin{matrix}\Delta^{-\frac{4}{3}}&{\rm ~~RD1},\\[0.1cm]1&{\rm ~~RD2}.\end{matrix}\right.\,
\label{eq:OmegaGWtoday_rough}
\eeqs

\begin{figure*}[htp]
  \centering
  \hspace*{-5mm}
  \subfigure{\includegraphics[scale=0.28]{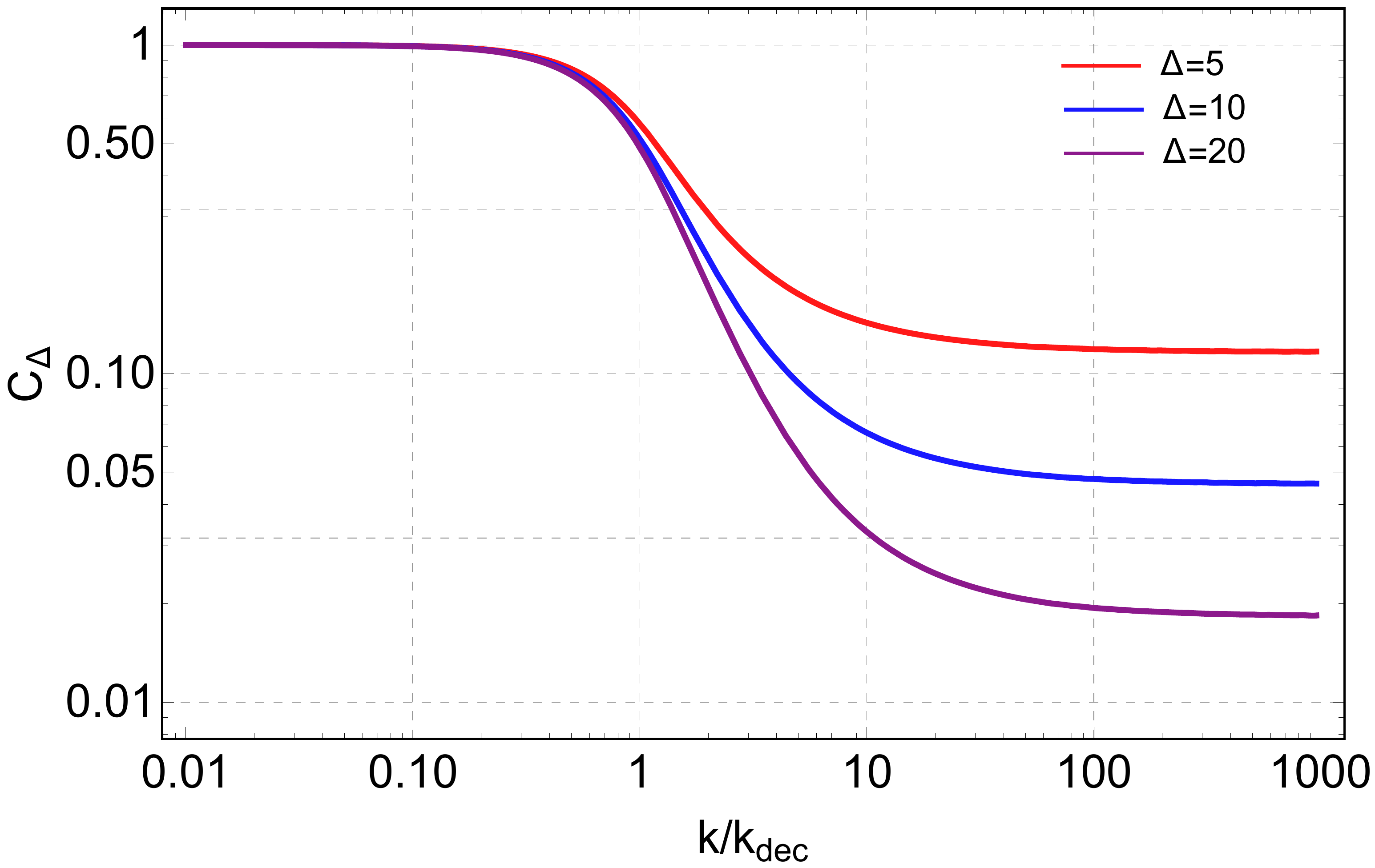}}
  \subfigure{\includegraphics[scale=0.28]{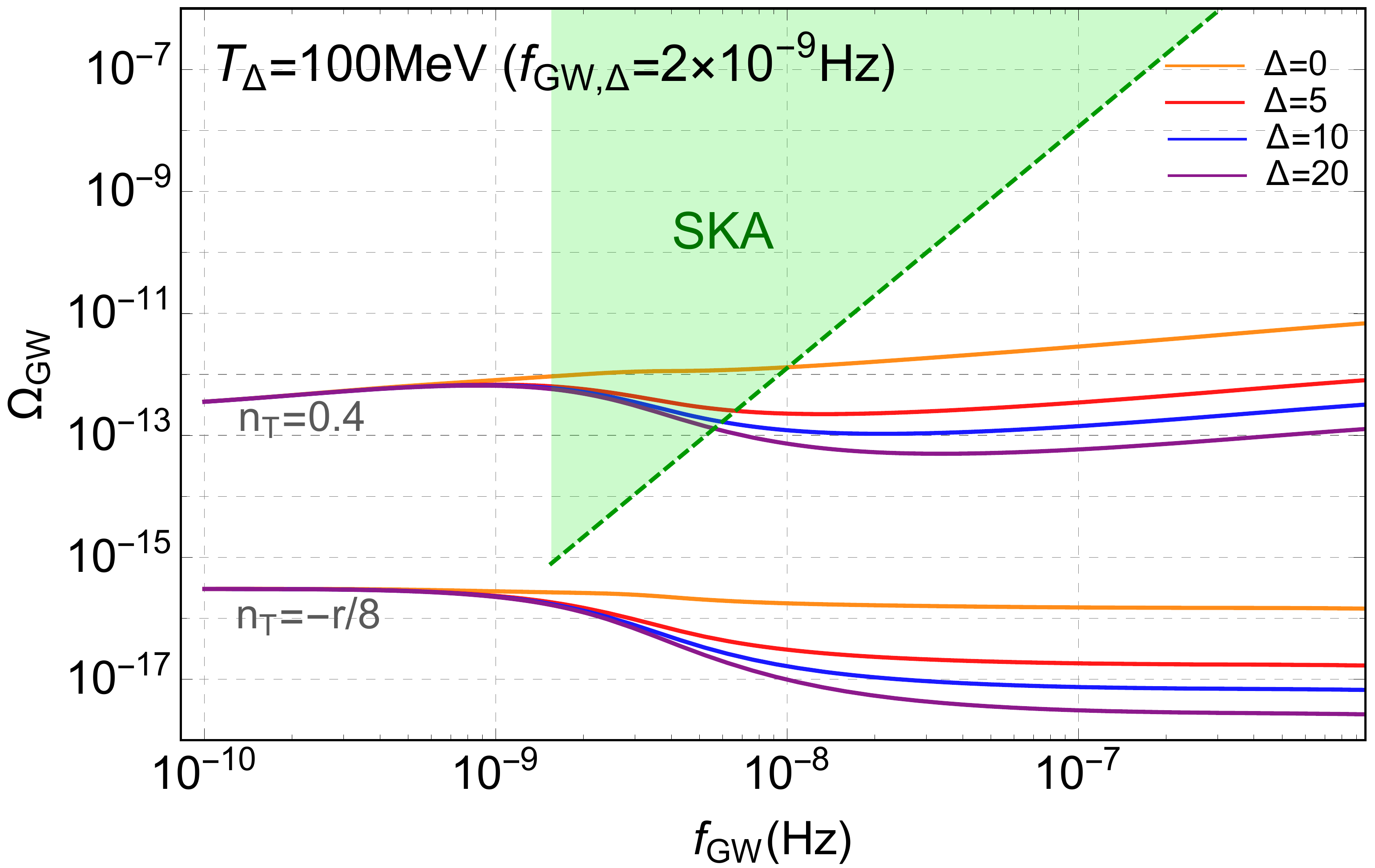}}
  \caption{{\bf Left panel}: The plot of $\mathcal{C}_{\Delta}(k)$ defined in Eq.~(\ref{eq:OmegaGWtoday}). Each suppression in $\mathcal{C}_{\Delta}(k)$ is due to the presence of the EMD era. {\bf Right panel}: The spectrum of pGWs in the Universe with $T_{\Delta}=100{\rm MeV}$ for $\mathcal{P}_{T}^{\rm prim}(k_{\rm CMB})=1.176\times10^{-10}$. The upper ($n_{T}=0.4$) and lower spectra ($n_{T}=-0.007$) result from the different choice of $n_{T}$. Suppression in red, blue and purple spectra is due to the specified dilution factor. The green dashed line is the sensitivity curve of SKA.}
  \vspace*{-1.5mm}
\label{fig:3}
\end{figure*}
\noindent

Here $\Omega_{\rm rad,0}\equiv\rho_{\rm rad,0}/\rho_{\rm cr,0}=4.2\times10^{-5}h^{-2}$ is the radiation energy fraction today, and $g_{*\rho}$ and $g_{*s}$ account for the relativistic degrees of freedom for the energy and entropy density, respectively.
We use $g_{*\rho,0}=3.383$ and $g_{*s,0}=3.931$ for their present values.
While Eq.~(\ref{eq:OmegaGWtoday_rough}) already tells us the rough behavior of the pGW spectrum at low and high wavenumbers, we parametrize it as\footnote{
Strictly speaking, this expression does not take into account the changes in the effective number of relativistic degrees of freedom during the finite period of horizon re-entry. We adopt this expression as an approximation, in order to separate the effect of relativistic degrees of freedom from that of the entropy production from the $X$ particle. Therefore, $\rho_{\rm rad}$ in Eq.~(\ref{eq:rhorad}) that appears later simply scales as $\propto a^{-4}$ in the absence of $X$ decay.
}

\beq
\Omega_{\rm GW,0}(k)=\frac{\Omega_{\rm rad,0}}{24}\left(\frac{g_{*\rho,k}}{g_{*\rho,0}}\right)\left(\frac{g_{*s,0}}{g_{*s,k}}\right)^{\!\frac{4}{3}}\mathcal{P}_{T}^{\rm prim}(k)\mathcal{C}_{\Delta}(k)\,,
\label{eq:OmegaGWtoday}
\eeq
in order to account for the wavenumbers re-entering the horizon during the entropy production.

To obtain $\mathcal{C}_{\Delta}(k)$, we explicitly calculate the time evolution of the pGW field. The final spectral shape depends on two parameters: the temperature $T_{\Delta}$ at which the entropy production occurs, and the dilution factor $\Delta\equiv S/\bar{S}$. Through $H(T_{\Delta})=\Gamma_{X}$, the first parameter can be exchanged with the decay rate of the heavy particle ($\Gamma_{X}$). We obtain the value of the second parameter once the value of $m_{3/2}$ is specified based on Fig.~\ref{fig:2}. Different combinations of the two parameters $(T_{\Delta},\Delta)$ give rise to different values of $\mathcal{C}_{\Delta}(k)$. This difference in $\mathcal{C}_{\Delta}(k)$ enables us to probe the early Universe history through $\Omega_{\rm GW}(f_{\rm GW})$ with $f_{\rm GW}=k/(2\pi a_{0})$. 

Before the heavy particle decays, the energy density of the Universe is mainly contributed from radiation ($\rho_{\rm rad}$) and the heavy particle $X$ ($\rho_{X}$). Thus the time evolution of the Hubble expansion rate in Eq.~(\ref{eq:hTEeqn}) is contributed by $\rho_{\rm rad}$ and $\rho_{X}$ through
\begin{eqnarray}
&&\dot{\rho}_{\rm rad}+4H\rho_{\rm rad}=\Gamma_{X}\rho_{X}\,,
\label{eq:rhorad}\cr\cr
&&\dot{\rho}_{X}+3H\rho_{X}=-\Gamma_{X}\rho_{X}\,,
\label{eq:rhoX}\cr\cr
&&H^{2}=\frac{\rho_{\rm rad}+\rho_{X}}{3M_{P}^{2}}\,.
\label{eq:Hubble}
\end{eqnarray}
When solved together with Eq.~(\ref{eq:Hubble}), Eq.~(\ref{eq:hTEeqn}) yields $h_{\bold{k}}^{+}(\tau)$ and $h_{\bold{k}}^{\times}(\tau)$ which in turn produce $\mathcal{C}_{\Delta}(k)$.

In the left panel of Fig.~\ref{fig:3}, we show $\mathcal{C}_{\Delta}(k)$ obtained by numerically solving Eq.~(\ref{eq:hTEeqn}) for different choices of the dilution factor $\Delta$. Each different color corresponds to the specified dilution factor. The suppression in $\mathcal{C}_{\Delta}(k)$ is observed for the mode $k\gtrsim0.1k_{\rm dec}$ where $k_{\rm dec}=a_{\rm dec}H_{\rm dec}$ is the mode that re-enters the horizon at the time when the decay of the heavy particle $X$ takes place $H_{\rm dec}=\Gamma_{X}$. Note that the value of the dilution factor is reflected in the ratio of the two plateaus which is equal to $\Delta^{-4/3}$. 

Concerning $g_{*\rho}$ and $g_{*s}$, since the MSSM sparticle mass spectrum depends on the details for how soft masses are generated, i.e. model-dependent, we do not study how GWs reflect the changes in these two for the temperature  $T\gtrsim T_{3/2,{\rm dec}}=\mathcal{O}(1){\rm TeV}$. Instead, we focus on those modes that re-enter the horizon for $T_{\rm MSSM}\lesssim T_{3/2,{\rm dec}}=\mathcal{O}(1){\rm TeV}$. Thus, for the temperature of the Universe of our interest, effectively only the SM particles and gravitinos exist. For the time evolution of $g_{*\rho}$ and $g_{*s}$ contributed by the SM particles, we shall refer to the tabulated data given in Ref.~\cite{Saikawa:2018rcs}. For the gravitino contribution, we evaluate the following:
\beq
g_{*\rho,3/2}(T)=2\left(\frac{T_{3/2}}{T}\right)^{\!4}\frac{15}{\pi^{4}}\int_{x_{3/2}}^{\infty}du\frac{u^{2}\sqrt{u^{2}-x_{3/2}^{2}}}{e^{u}+1}\,,
\label{eq:grho32}
\eeq
\beqs
g_{*s,3/2}(T)&=&2\left(\frac{T_{3/2}}{T}\right)^{\!3}\frac{15}{\pi^{4}}\cr\cr&\times&\int_{x_{3/2}}^{\infty}du\frac{(u^{2}-x_{3/2}^{2}/4)\sqrt{u^{2}-x_{3/2}^{2}}}{e^{u}+1}\,,
\label{eq:gs32}
\eeqs
where we defined $x_{3/2}=m_{3/2}/T$ and we assume the chemical potential of the gravitino to be negligible. Adding Eqs.~(\ref{eq:grho32}) and (\ref{eq:gs32}) to $g_{*\rho}$ and $g_{*s}$ for the SM, we obtain the final $g_{*\rho}$ and $g_{*s}$.

We show in the right panel of Fig.~\ref{fig:3} the pGW spectrum $\Omega_{\rm GW}(f_{\rm GW})$ computed from Eq.~(\ref{eq:OmegaGWtoday}). The pGW frequency and the temperature $T_{\rm hc}$ when the relevant mode re-enters the horizon are related via
\beqs
f_{\rm GW}&=&\frac{k}{2\pi a_{0}}\cr\cr
&\simeq&2.65{\rm Hz}\left(\frac{g_{*\rho,k}}{106.75}\right)^{\!\frac{1}{2}}\left(\frac{g_{*s,k}}{106.75}\right)^{\!-\frac{1}{3}}\left(\frac{T_{\rm hc}}{10^{8}{\rm GeV}}\right)\,.\nonumber\\
\label{eq:fTrelation}
\eeqs
To see how $\Omega_{\rm GW}$ depends on the inflation model, we assume the standard single-field slow-roll inflation of $n_{T}=-r_{\rm max}/8=-0.007$ with $r_{\rm max}$ the maximum allowed tensor-to-scalar ratio, and a non-minimal inflation model with $n_{T}=0.4$~\cite{Cook:2011hg,Barnaby:2011qe,Anber:2012du,Domcke:2016bkh,Jimenez:2017cdr,Papageorgiou:2019ecb}.\footnote{As for the choice of $n_{T}=0.4$, we refer to Ref.~\cite{DEramo:2019tit} where $n_{T}=0.4$ was chosen to assess the maximal reach of future GW experiments and the explanation for consistency with CMB observation~\cite{Akrami:2018odb}, constraints from LIGO and Virgo~\cite{TheLIGOScientific:2016dpb,LIGOScientific:2019vic}, and BBN, LIGO, and pulsars~\cite{Kuroyanagi:2014nba} was made.} As an example, we take $T_{\Delta}=100{\rm MeV}$ with $\Delta=0,5,10$ and $20$. 

We observe that $\Omega_{\rm GW}$ with $\Delta\neq0$ is characterized by the suppression at $f_{\rm GW,\Delta}$ corresponding to $T_{\Delta}$ (related via Eq.~(\ref{eq:fTrelation})), which can be used to investigate whether the Universe went through the entropy production during the RD era. Larger $\Delta$ induces a greater suppression in $\Omega_{\rm GW}$ for $f_{\rm GW}>f_{\rm GW,\Delta}$. The underlying reason for the suppression is the presence of the EMD era where $\Omega_{\rm GW}\propto a^{-1}$. Because $\Omega_{\rm GW}\propto a^{0}$ holds true during the RD2 era, given the almost scale-invariant initial spectra on the super-horizon scale, dilution of $\Omega_{\rm GW}$ during the EMD era causes the suppression in the spectrum for the modes $k\gtrsim k_{\rm dec}$. Although only the case with $T_{\Delta}=100{\rm MeV}$ is shown in the right panel of Fig.~\ref{fig:3}, for other choices of $T_{\Delta}$ we expect a similar spectral shape with suppression at the relevant $f_{\rm GW,\Delta}$.

It should be noted that $T_{\Delta}$ is limited to the range $10{\rm MeV}\lesssim T_{\Delta}\lesssim\mathcal{O}(1){\rm TeV}$. The lower bound is to ensure the successful BBN~\cite{Kawasaki:2004qu,Hasegawa:2019jsa} (see also references therein). On the other hand, as discussed in Sec.~\ref{sec:sec2}, $T_{3/2,{\rm dec}}$ can be at most $\mathcal{O}(10){\rm TeV}$ since $m_{\tilde{g}}\lesssim\mathcal{O}(10){\rm TeV}$. Since the entropy production should happen after gravitinos decouple from the thermal bath, we conclude that $T_{\Delta}\lesssim\mathcal{O}(1){\rm TeV}$ should be the case. Hence, for a PeV SUSY-breaking scenario, we can make a quite solid argument that the suppression of $\Omega_{\rm GW}$ due to the entropy production should occur at $f_{\rm GW,\Delta}=\mathcal{O}(10^{-10}){\rm Hz}-\mathcal{O}(10^{-5}){\rm Hz}$, irrespective of the inflation model assumed.

In the right panel of Fig.~\ref{fig:3}, given the green shaded region showing the parameter space potentially probed by SKA~\cite{Carilli:2004nx,Janssen:2014dka,Bull:2018lat}, we notice that SKA has a chance to see the suppression directly for $T_{\Delta}=\mathcal{O}(100){\rm MeV}$ when $n_{T}=0.4$. However, for other cases with much different values of $T_{\Delta}$ or with $n_{T}=-r/8$, direct observation of the suppression in $\Omega_{\rm GW}$ with SKA may not be possible. In this regard, the detection of CMB B-mode polarization in the future can play a critical role, even if the relevant frequency is too low to directly probe $f_{\rm GW,\Delta}$ of our interest. 
A careful comparison between the amount of pGWs observed at the CMB scales and at high frequencies by space-based interferometers such as LISA~\cite{eLISA:2013xep,LISA:2017pwj}, DECIGO~\cite{Seto:2001qf,Kawamura:2006up}, Taiji~\cite{Hu:2017mde}, TianQin~\cite{TianQin:2015yph,TianQin:2020hid}, and BBO~\cite{Crowder:2005nr,Corbin:2005ny,Harry:2006fi}, may indirectly reveal the presence of the suppression in $\Omega_{\rm GW}$.
Although this way of indirect investigation cannot pin down $T_{\Delta}$, the entropy production $\Delta$ can at least be determined.
Therefore, the synergy between CMB B-mode polarization surveys and future space-based interferometers still renders PeV SUSY-breaking scenarios testable even if future pulsar timing arrays (such as IPTA~\cite{Hobbs:2009yy,Manchester:2013ndt,Verbiest:2016vem} and SKA) are not sensitive enough to probe the entropy production at $10{\rm MeV}\lesssim T_{\Delta}\lesssim\mathcal{O}(1){\rm TeV}$.\footnote{Note that at the GW frequency range relevant to DECIGO and BBO, some or all of the MSSM particles are relativistic. Thus, when probing the suppression of $\Omega_{\rm GW}$ due to the entropy production using DECIGO and BBO, the additional suppression caused by the larger $g_{*\rho}(T_{\rm in})$ and $g_{*s}(T_{\rm in})$ in Eq.~(\ref{eq:OmegaGWtoday}) should be taken into account.}

We end this section by commenting on effects that sub-keV gravitinos can have on the pGW spectrum as free-streaming dark radiation (DR)~\cite{1982ApJ...257..456V,Rebhan:1994zw,Weinberg:2003ur}.\footnote{
See \cite{Hook:2020phx} for another interesting effect of free-streaming radiations on GWs. It is an effect on the IR ``causal" part of the spectrum in late time GW production.
}
Once gravitinos decouple from the MSSM thermal bath at $T_{3/2,{\rm dec}}=\mathcal{O}(1){\rm TeV}$, they start to behave as DR. Without a significant late time entropy production after decoupling, $\Delta N_{\rm eff}$ contributed by gravitinos (say $N_{3/2}$) is given by
\beqs
N_{3/2}&=&\left(\frac{T_{3/2}}{T_{\nu}}\right)^{\!4}=\left(\frac{g_{*s}(T_{\nu,{\rm dec}})}{g_{*s}(T_{3/2,{\rm dec}})}\right)^{\!\frac{4}{3}}\cr\cr&\simeq&0.017\left(\frac{g_{*s}(T_{3/2,{\rm dec}})}{230}\right)^{\!-\frac{4}{3}}\,,
\label{eq:N32}
\eeqs
where $T_{\nu,{\rm dec}}$ is the neutrino decoupling temperature and $g_{*s}(T_{\nu,{\rm dec}})=10.75$ is the effective number of degrees of freedom for the entropy density evaluated at $T_{\nu,{\rm dec}}$. If DR makes a significant contribution to $\Delta N_{\rm eff}$, there can be an overall enhancement of the pGW spectrum (due to the change in the expansion rate and that in the matter-radiation equality) and a suppression feature from the non-zero anisotropic stress induced by the free-streaming of DR~\cite{Boyle:2005se,Jinno:2012xb}. However, since $N_{3/2}$ is too small for the scenarios discussed in the present paper, these effects from $N_{3/2}$ are too small to be visible in the right panel of Fig.~\ref{fig:3}.


\section{Conclusion}
\label{sec:conclusion}
In this work, we demonstrated the necessity of the late time entropy production prior to BBN era for the Universe with PeV scale SUSY-breaking. Our argument is based on the observation that the theoretically predicted relic abundance of gravitino WDM with $m_{3/2}\in[100{\rm eV},1{\rm keV}]$ is too large to be consistent with the observed number of satellite galaxies $N_{\rm sat}$ in the Milky Way, provided that there is no energy dilution mechanism. By requiring $N_{\rm sat}\gtrsim63$ in the current Universe in which sub-keV gravitino WDM is responsible for a fraction of the DM population, we quantified the minimum amount of the dilution factor $\Delta_{\rm min}$ necessary for observational consistency. 

Based on this result, we proposed using the pGW spectrum $\Omega_{\rm GW}$ to investigate the presence of an early matter-dominated era and entropy production prior to the BBN era. The degree and the characteristic frequency of the suppression in the pGW spectrum, once observed by future pulsar timing arrays such as SKA, give us the information on the entropy production $\Delta$ and the temperature $T_{\Delta}$ when it occurs. Interestingly, the suppression is expected at $\mathcal{O}(10^{-10}){\rm Hz}\lesssim f_{\rm GW}\lesssim\mathcal{O}(10^{-5}){\rm Hz}$ in PeV scale SUSY-breaking scenarios, making these scenarios testable via pGW searches. Even if the pGW abundance is too small to be probed by SKA, the synergy between future CMB B-mode experiments and space-based interferometers can alternatively probe the suppression in the pGW spectrum induced by $\Delta\neq0$. If the absence of any suppression feature in the pGW spectrum is established by the comparison between the two frequency regimes $f_{\rm GW}<\mathcal{O}(10^{-10}){\rm Hz}$ and $f_{\rm GW}>\mathcal{O}(10^{-5}){\rm Hz}$, our study immediately rules out PeV scale SUSY-breaking scenarios.

We have concentrated on the case of $m_{3/2}=\mathcal{O}(100){\rm eV}$ in this paper. However, we stress that our method of using pGWs can be extended to test the low scale SUSY with larger mass range of the 
gravitino such as $1{\rm keV}-1{\rm GeV}$ which is predicted in gauge mediation 
models~\cite{Hamaguchi:2014sea,Choi:2020wdq}.


\begin{acknowledgments}
The work of R.J. is supported by Grants-in-Aid for JSPS Overseas Research Fellow (No. 201960698). This work is supported by the Deutsche Forschungsgemeinschaft 
under Germany's Excellence Strategy -- EXC 2121 ,,Quantum Universe`` -- 390833306.
T. T. Y. is supported in part by the China Grant for Talent Scientific Start-Up Project and the JSPS Grant-in-Aid for Scientific Research No. 16H02176, No. 17H02878, and No. 19H05810 and by World Premier International Research Center Initiative (WPI Initiative), MEXT, Japan. 

\end{acknowledgments}


\bibliography{main}

\begin{thebibliography}{87}%
\makeatletter
\providecommand \@ifxundefined [1]{%
 \@ifx{#1\undefined}
}%
\providecommand \@ifnum [1]{%
 \ifnum #1\expandafter \@firstoftwo
 \else \expandafter \@secondoftwo
 \fi
}%
\providecommand \@ifx [1]{%
 \ifx #1\expandafter \@firstoftwo
 \else \expandafter \@secondoftwo
 \fi
}%
\providecommand \natexlab [1]{#1}%
\providecommand \enquote  [1]{``#1''}%
\providecommand \bibnamefont  [1]{#1}%
\providecommand \bibfnamefont [1]{#1}%
\providecommand \citenamefont [1]{#1}%
\providecommand \href@noop [0]{\@secondoftwo}%
\providecommand \href [0]{\begingroup \@sanitize@url \@href}%
\providecommand \@href[1]{\@@startlink{#1}\@@href}%
\providecommand \@@href[1]{\endgroup#1\@@endlink}%
\providecommand \@sanitize@url [0]{\catcode `\\12\catcode `\$12\catcode
  `\&12\catcode `\#12\catcode `\^12\catcode `\_12\catcode `\%12\relax}%
\providecommand \@@startlink[1]{}%
\providecommand \@@endlink[0]{}%
\providecommand \url  [0]{\begingroup\@sanitize@url \@url }%
\providecommand \@url [1]{\endgroup\@href {#1}{\urlprefix }}%
\providecommand \urlprefix  [0]{URL }%
\providecommand \Eprint [0]{\href }%
\providecommand \doibase [0]{http://dx.doi.org/}%
\providecommand \selectlanguage [0]{\@gobble}%
\providecommand \bibinfo  [0]{\@secondoftwo}%
\providecommand \bibfield  [0]{\@secondoftwo}%
\providecommand \translation [1]{[#1]}%
\providecommand \BibitemOpen [0]{}%
\providecommand \bibitemStop [0]{}%
\providecommand \bibitemNoStop [0]{.\EOS\space}%
\providecommand \EOS [0]{\spacefactor3000\relax}%
\providecommand \BibitemShut  [1]{\csname bibitem#1\endcsname}%
\let\auto@bib@innerbib\@empty
\bibitem [{\citenamefont {Yanagida}\ \emph {et~al.}(2012)\citenamefont
  {Yanagida}, \citenamefont {Yokozaki},\ and\ \citenamefont
  {Yonekura}}]{Yanagida:2012ef}%
  \BibitemOpen
  \bibfield  {author} {\bibinfo {author} {\bibfnamefont {T.~T.}\ \bibnamefont
  {Yanagida}}, \bibinfo {author} {\bibfnamefont {N.}~\bibnamefont {Yokozaki}},
  \ and\ \bibinfo {author} {\bibfnamefont {K.}~\bibnamefont {Yonekura}},\
  }\href {\doibase 10.1007/JHEP10(2012)017} {\bibfield  {journal} {\bibinfo
  {journal} {JHEP}\ }\textbf {\bibinfo {volume} {10}},\ \bibinfo {pages} {017}
  (\bibinfo {year} {2012})},\ \Eprint {http://arxiv.org/abs/1206.6589}
  {arXiv:1206.6589 [hep-ph]} \BibitemShut {NoStop}%
\bibitem [{\citenamefont {Moroi}\ \emph {et~al.}(1993)\citenamefont {Moroi},
  \citenamefont {Murayama},\ and\ \citenamefont {Yamaguchi}}]{Moroi:1993mb}%
  \BibitemOpen
  \bibfield  {author} {\bibinfo {author} {\bibfnamefont {T.}~\bibnamefont
  {Moroi}}, \bibinfo {author} {\bibfnamefont {H.}~\bibnamefont {Murayama}}, \
  and\ \bibinfo {author} {\bibfnamefont {M.}~\bibnamefont {Yamaguchi}},\ }\href
  {\doibase 10.1016/0370-2693(93)91434-O} {\bibfield  {journal} {\bibinfo
  {journal} {Phys. Lett. B}\ }\textbf {\bibinfo {volume} {303}},\ \bibinfo
  {pages} {289} (\bibinfo {year} {1993})}\BibitemShut {NoStop}%
\bibitem [{\citenamefont {Bowman}\ \emph {et~al.}(2018)\citenamefont {Bowman},
  \citenamefont {Rogers}, \citenamefont {Monsalve}, \citenamefont {Mozdzen},\
  and\ \citenamefont {Mahesh}}]{Bowman:2018yin}%
  \BibitemOpen
  \bibfield  {author} {\bibinfo {author} {\bibfnamefont {J.~D.}\ \bibnamefont
  {Bowman}}, \bibinfo {author} {\bibfnamefont {A.~E.~E.}\ \bibnamefont
  {Rogers}}, \bibinfo {author} {\bibfnamefont {R.~A.}\ \bibnamefont
  {Monsalve}}, \bibinfo {author} {\bibfnamefont {T.~J.}\ \bibnamefont
  {Mozdzen}}, \ and\ \bibinfo {author} {\bibfnamefont {N.}~\bibnamefont
  {Mahesh}},\ }\href {\doibase 10.1038/nature25792} {\bibfield  {journal}
  {\bibinfo  {journal} {Nature}\ }\textbf {\bibinfo {volume} {555}},\ \bibinfo
  {pages} {67} (\bibinfo {year} {2018})},\ \Eprint
  {http://arxiv.org/abs/1810.05912} {arXiv:1810.05912 [astro-ph.CO]}
  \BibitemShut {NoStop}%
\bibitem [{\citenamefont {Ir\v~si\v c}\ \emph {et~al.}(2017)\citenamefont
  {Ir\v~si\v c} \emph {et~al.}}]{Irsic:2017ixq}%
  \BibitemOpen
  \bibfield  {author} {\bibinfo {author} {\bibfnamefont {V.}~\bibnamefont
  {Ir\v~si\v c}} \emph {et~al.},\ }\href {\doibase 10.1103/PhysRevD.96.023522}
  {\bibfield  {journal} {\bibinfo  {journal} {Phys. Rev. D}\ }\textbf {\bibinfo
  {volume} {96}},\ \bibinfo {pages} {023522} (\bibinfo {year} {2017})},\
  \Eprint {http://arxiv.org/abs/1702.01764} {arXiv:1702.01764 [astro-ph.CO]}
  \BibitemShut {NoStop}%
\bibitem [{\citenamefont {Schneider}(2018)}]{Schneider:2018xba}%
  \BibitemOpen
  \bibfield  {author} {\bibinfo {author} {\bibfnamefont {A.}~\bibnamefont
  {Schneider}},\ }\href {\doibase 10.1103/PhysRevD.98.063021} {\bibfield
  {journal} {\bibinfo  {journal} {Phys. Rev. D}\ }\textbf {\bibinfo {volume}
  {98}},\ \bibinfo {pages} {063021} (\bibinfo {year} {2018})},\ \Eprint
  {http://arxiv.org/abs/1805.00021} {arXiv:1805.00021 [astro-ph.CO]}
  \BibitemShut {NoStop}%
\bibitem [{\citenamefont {Lopez-Honorez}\ \emph {et~al.}(2019)\citenamefont
  {Lopez-Honorez}, \citenamefont {Mena},\ and\ \citenamefont
  {Villanueva-Domingo}}]{Lopez-Honorez:2018ipk}%
  \BibitemOpen
  \bibfield  {author} {\bibinfo {author} {\bibfnamefont {L.}~\bibnamefont
  {Lopez-Honorez}}, \bibinfo {author} {\bibfnamefont {O.}~\bibnamefont {Mena}},
  \ and\ \bibinfo {author} {\bibfnamefont {P.}~\bibnamefont
  {Villanueva-Domingo}},\ }\href {\doibase 10.1103/PhysRevD.99.023522}
  {\bibfield  {journal} {\bibinfo  {journal} {Phys. Rev. D}\ }\textbf {\bibinfo
  {volume} {99}},\ \bibinfo {pages} {023522} (\bibinfo {year} {2019})},\
  \Eprint {http://arxiv.org/abs/1811.02716} {arXiv:1811.02716 [astro-ph.CO]}
  \BibitemShut {NoStop}%
\bibitem [{\citenamefont {Garzilli}\ \emph {et~al.}(2019)\citenamefont
  {Garzilli}, \citenamefont {Ruchayskiy}, \citenamefont {Magalich},\ and\
  \citenamefont {Boyarsky}}]{Garzilli:2019qki}%
  \BibitemOpen
  \bibfield  {author} {\bibinfo {author} {\bibfnamefont {A.}~\bibnamefont
  {Garzilli}}, \bibinfo {author} {\bibfnamefont {O.}~\bibnamefont
  {Ruchayskiy}}, \bibinfo {author} {\bibfnamefont {A.}~\bibnamefont
  {Magalich}}, \ and\ \bibinfo {author} {\bibfnamefont {A.}~\bibnamefont
  {Boyarsky}},\ }\href@noop {} {\  (\bibinfo {year} {2019})},\ \Eprint
  {http://arxiv.org/abs/1912.09397} {arXiv:1912.09397 [astro-ph.CO]}
  \BibitemShut {NoStop}%
\bibitem [{\citenamefont {Grishchuk}(1974)}]{Grishchuk:1974ny}%
  \BibitemOpen
  \bibfield  {author} {\bibinfo {author} {\bibfnamefont {L.~P.}\ \bibnamefont
  {Grishchuk}},\ }\href@noop {} {\bibfield  {journal} {\bibinfo  {journal} {Zh.
  Eksp. Teor. Fiz.}\ }\textbf {\bibinfo {volume} {67}},\ \bibinfo {pages} {825}
  (\bibinfo {year} {1974})}\BibitemShut {NoStop}%
\bibitem [{\citenamefont {Starobinsky}(1979)}]{Starobinsky:1979ty}%
  \BibitemOpen
  \bibfield  {author} {\bibinfo {author} {\bibfnamefont {A.~A.}\ \bibnamefont
  {Starobinsky}},\ }\href@noop {} {\bibfield  {journal} {\bibinfo  {journal}
  {JETP Lett.}\ }\textbf {\bibinfo {volume} {30}},\ \bibinfo {pages} {682}
  (\bibinfo {year} {1979})}\BibitemShut {NoStop}%
\bibitem [{\citenamefont {Turner}\ \emph {et~al.}(1993)\citenamefont {Turner},
  \citenamefont {White},\ and\ \citenamefont {Lidsey}}]{Turner:1993vb}%
  \BibitemOpen
  \bibfield  {author} {\bibinfo {author} {\bibfnamefont {M.~S.}\ \bibnamefont
  {Turner}}, \bibinfo {author} {\bibfnamefont {M.~J.}\ \bibnamefont {White}}, \
  and\ \bibinfo {author} {\bibfnamefont {J.~E.}\ \bibnamefont {Lidsey}},\
  }\href {\doibase 10.1103/PhysRevD.48.4613} {\bibfield  {journal} {\bibinfo
  {journal} {Phys. Rev. D}\ }\textbf {\bibinfo {volume} {48}},\ \bibinfo
  {pages} {4613} (\bibinfo {year} {1993})},\ \Eprint
  {http://arxiv.org/abs/astro-ph/9306029} {arXiv:astro-ph/9306029} \BibitemShut
  {NoStop}%
\bibitem [{\citenamefont {Turner}(1997)}]{Turner:1996ck}%
  \BibitemOpen
  \bibfield  {author} {\bibinfo {author} {\bibfnamefont {M.~S.}\ \bibnamefont
  {Turner}},\ }\href {\doibase 10.1103/PhysRevD.55.R435} {\bibfield  {journal}
  {\bibinfo  {journal} {Phys. Rev. D}\ }\textbf {\bibinfo {volume} {55}},\
  \bibinfo {pages} {R435} (\bibinfo {year} {1997})},\ \Eprint
  {http://arxiv.org/abs/astro-ph/9607066} {arXiv:astro-ph/9607066} \BibitemShut
  {NoStop}%
\bibitem [{\citenamefont {Smith}\ \emph {et~al.}(2006)\citenamefont {Smith},
  \citenamefont {Kamionkowski},\ and\ \citenamefont {Cooray}}]{Smith:2005mm}%
  \BibitemOpen
  \bibfield  {author} {\bibinfo {author} {\bibfnamefont {T.~L.}\ \bibnamefont
  {Smith}}, \bibinfo {author} {\bibfnamefont {M.}~\bibnamefont {Kamionkowski}},
  \ and\ \bibinfo {author} {\bibfnamefont {A.}~\bibnamefont {Cooray}},\ }\href
  {\doibase 10.1103/PhysRevD.73.023504} {\bibfield  {journal} {\bibinfo
  {journal} {Phys. Rev. D}\ }\textbf {\bibinfo {volume} {73}},\ \bibinfo
  {pages} {023504} (\bibinfo {year} {2006})},\ \Eprint
  {http://arxiv.org/abs/astro-ph/0506422} {arXiv:astro-ph/0506422} \BibitemShut
  {NoStop}%
\bibitem [{\citenamefont {Hisano}\ \emph {et~al.}(2008)\citenamefont {Hisano},
  \citenamefont {Nagai}, \citenamefont {Sugiyama},\ and\ \citenamefont
  {Yanagida}}]{Hisano:2008sy}%
  \BibitemOpen
  \bibfield  {author} {\bibinfo {author} {\bibfnamefont {J.}~\bibnamefont
  {Hisano}}, \bibinfo {author} {\bibfnamefont {M.}~\bibnamefont {Nagai}},
  \bibinfo {author} {\bibfnamefont {S.}~\bibnamefont {Sugiyama}}, \ and\
  \bibinfo {author} {\bibfnamefont {T.~T.}\ \bibnamefont {Yanagida}},\ }\href
  {\doibase 10.1016/j.physletb.2008.06.045} {\bibfield  {journal} {\bibinfo
  {journal} {Phys. Lett. B}\ }\textbf {\bibinfo {volume} {665}},\ \bibinfo
  {pages} {237} (\bibinfo {year} {2008})},\ \Eprint
  {http://arxiv.org/abs/0804.2957} {arXiv:0804.2957 [hep-ph]} \BibitemShut
  {NoStop}%
\bibitem [{\citenamefont {Viel}\ \emph {et~al.}(2005)\citenamefont {Viel},
  \citenamefont {Lesgourgues}, \citenamefont {Haehnelt}, \citenamefont
  {Matarrese},\ and\ \citenamefont {Riotto}}]{Viel:2005qj}%
  \BibitemOpen
  \bibfield  {author} {\bibinfo {author} {\bibfnamefont {M.}~\bibnamefont
  {Viel}}, \bibinfo {author} {\bibfnamefont {J.}~\bibnamefont {Lesgourgues}},
  \bibinfo {author} {\bibfnamefont {M.~G.}\ \bibnamefont {Haehnelt}}, \bibinfo
  {author} {\bibfnamefont {S.}~\bibnamefont {Matarrese}}, \ and\ \bibinfo
  {author} {\bibfnamefont {A.}~\bibnamefont {Riotto}},\ }\href {\doibase
  10.1103/PhysRevD.71.063534} {\bibfield  {journal} {\bibinfo  {journal} {Phys.
  Rev. D}\ }\textbf {\bibinfo {volume} {71}},\ \bibinfo {pages} {063534}
  (\bibinfo {year} {2005})},\ \Eprint {http://arxiv.org/abs/astro-ph/0501562}
  {arXiv:astro-ph/0501562} \BibitemShut {NoStop}%
\bibitem [{\citenamefont {Fujii}\ and\ \citenamefont
  {Yanagida}(2002)}]{Fujii:2002fv}%
  \BibitemOpen
  \bibfield  {author} {\bibinfo {author} {\bibfnamefont {M.}~\bibnamefont
  {Fujii}}\ and\ \bibinfo {author} {\bibfnamefont {T.}~\bibnamefont
  {Yanagida}},\ }\href {\doibase 10.1016/S0370-2693(02)02958-1} {\bibfield
  {journal} {\bibinfo  {journal} {Phys. Lett. B}\ }\textbf {\bibinfo {volume}
  {549}},\ \bibinfo {pages} {273} (\bibinfo {year} {2002})},\ \Eprint
  {http://arxiv.org/abs/hep-ph/0208191} {arXiv:hep-ph/0208191} \BibitemShut
  {NoStop}%
\bibitem [{\citenamefont {Ibe}\ \emph {et~al.}(2011)\citenamefont {Ibe},
  \citenamefont {Sato}, \citenamefont {Yanagida},\ and\ \citenamefont
  {Yonekura}}]{Ibe:2010ym}%
  \BibitemOpen
  \bibfield  {author} {\bibinfo {author} {\bibfnamefont {M.}~\bibnamefont
  {Ibe}}, \bibinfo {author} {\bibfnamefont {R.}~\bibnamefont {Sato}}, \bibinfo
  {author} {\bibfnamefont {T.~T.}\ \bibnamefont {Yanagida}}, \ and\ \bibinfo
  {author} {\bibfnamefont {K.}~\bibnamefont {Yonekura}},\ }\href {\doibase
  10.1007/JHEP04(2011)077} {\bibfield  {journal} {\bibinfo  {journal} {JHEP}\
  }\textbf {\bibinfo {volume} {04}},\ \bibinfo {pages} {077} (\bibinfo {year}
  {2011})},\ \Eprint {http://arxiv.org/abs/1012.5466} {arXiv:1012.5466
  [hep-ph]} \BibitemShut {NoStop}%
\bibitem [{\citenamefont {Choi}\ and\ \citenamefont
  {Yanagida}(2021)}]{Choi:2021uhy}%
  \BibitemOpen
  \bibfield  {author} {\bibinfo {author} {\bibfnamefont {G.}~\bibnamefont
  {Choi}}\ and\ \bibinfo {author} {\bibfnamefont {T.~T.}\ \bibnamefont
  {Yanagida}},\ }\href@noop {} {\  (\bibinfo {year} {2021})},\ \Eprint
  {http://arxiv.org/abs/2104.02958} {arXiv:2104.02958 [hep-ph]} \BibitemShut
  {NoStop}%
\bibitem [{\citenamefont {Boyarsky}\ \emph {et~al.}(2009)\citenamefont
  {Boyarsky}, \citenamefont {Lesgourgues}, \citenamefont {Ruchayskiy},\ and\
  \citenamefont {Viel}}]{Boyarsky:2008xj}%
  \BibitemOpen
  \bibfield  {author} {\bibinfo {author} {\bibfnamefont {A.}~\bibnamefont
  {Boyarsky}}, \bibinfo {author} {\bibfnamefont {J.}~\bibnamefont
  {Lesgourgues}}, \bibinfo {author} {\bibfnamefont {O.}~\bibnamefont
  {Ruchayskiy}}, \ and\ \bibinfo {author} {\bibfnamefont {M.}~\bibnamefont
  {Viel}},\ }\href {\doibase 10.1088/1475-7516/2009/05/012} {\bibfield
  {journal} {\bibinfo  {journal} {JCAP}\ }\textbf {\bibinfo {volume} {05}},\
  \bibinfo {pages} {012} (\bibinfo {year} {2009})},\ \Eprint
  {http://arxiv.org/abs/0812.0010} {arXiv:0812.0010 [astro-ph]} \BibitemShut
  {NoStop}%
\bibitem [{\citenamefont {Blas}\ \emph {et~al.}(2011)\citenamefont {Blas},
  \citenamefont {Lesgourgues},\ and\ \citenamefont {Tram}}]{Blas:2011rf}%
  \BibitemOpen
  \bibfield  {author} {\bibinfo {author} {\bibfnamefont {D.}~\bibnamefont
  {Blas}}, \bibinfo {author} {\bibfnamefont {J.}~\bibnamefont {Lesgourgues}}, \
  and\ \bibinfo {author} {\bibfnamefont {T.}~\bibnamefont {Tram}},\ }\href
  {\doibase 10.1088/1475-7516/2011/07/034} {\bibfield  {journal} {\bibinfo
  {journal} {JCAP}\ }\textbf {\bibinfo {volume} {1107}},\ \bibinfo {pages}
  {034} (\bibinfo {year} {2011})},\ \Eprint {http://arxiv.org/abs/1104.2933}
  {arXiv:1104.2933 [astro-ph.CO]} \BibitemShut {NoStop}%
\bibitem [{\citenamefont {Polisensky}\ and\ \citenamefont
  {Ricotti}(2011)}]{Polisensky:2010rw}%
  \BibitemOpen
  \bibfield  {author} {\bibinfo {author} {\bibfnamefont {E.}~\bibnamefont
  {Polisensky}}\ and\ \bibinfo {author} {\bibfnamefont {M.}~\bibnamefont
  {Ricotti}},\ }\href {\doibase 10.1103/PhysRevD.83.043506} {\bibfield
  {journal} {\bibinfo  {journal} {Phys. Rev. D}\ }\textbf {\bibinfo {volume}
  {83}},\ \bibinfo {pages} {043506} (\bibinfo {year} {2011})},\ \Eprint
  {http://arxiv.org/abs/1004.1459} {arXiv:1004.1459 [astro-ph.CO]} \BibitemShut
  {NoStop}%
\bibitem [{\citenamefont {Giocoli}\ \emph {et~al.}(2008)\citenamefont
  {Giocoli}, \citenamefont {Pieri},\ and\ \citenamefont
  {Tormen}}]{Giocoli:2007gf}%
  \BibitemOpen
  \bibfield  {author} {\bibinfo {author} {\bibfnamefont {C.}~\bibnamefont
  {Giocoli}}, \bibinfo {author} {\bibfnamefont {L.}~\bibnamefont {Pieri}}, \
  and\ \bibinfo {author} {\bibfnamefont {G.}~\bibnamefont {Tormen}},\ }\href
  {\doibase 10.1111/j.1365-2966.2008.13283.x} {\bibfield  {journal} {\bibinfo
  {journal} {Mon. Not. Roy. Astron. Soc.}\ }\textbf {\bibinfo {volume} {387}},\
  \bibinfo {pages} {689} (\bibinfo {year} {2008})},\ \Eprint
  {http://arxiv.org/abs/0712.1476} {arXiv:0712.1476 [astro-ph]} \BibitemShut
  {NoStop}%
\bibitem [{\citenamefont {Maccio}\ and\ \citenamefont
  {Fontanot}(2010)}]{Maccio:2009isa}%
  \BibitemOpen
  \bibfield  {author} {\bibinfo {author} {\bibfnamefont {A.~V.}\ \bibnamefont
  {Maccio}}\ and\ \bibinfo {author} {\bibfnamefont {F.}~\bibnamefont
  {Fontanot}},\ }\href {\doibase 10.1111/j.1745-3933.2010.00825.x} {\bibfield
  {journal} {\bibinfo  {journal} {Mon. Not. Roy. Astron. Soc.}\ }\textbf
  {\bibinfo {volume} {404}},\ \bibinfo {pages} {16} (\bibinfo {year} {2010})},\
  \Eprint {http://arxiv.org/abs/0910.2460} {arXiv:0910.2460 [astro-ph.CO]}
  \BibitemShut {NoStop}%
\bibitem [{\citenamefont {Horiuchi}\ \emph {et~al.}(2014)\citenamefont
  {Horiuchi}, \citenamefont {Humphrey}, \citenamefont {Onorbe}, \citenamefont
  {Abazajian}, \citenamefont {Kaplinghat},\ and\ \citenamefont
  {Garrison-Kimmel}}]{Horiuchi:2013noa}%
  \BibitemOpen
  \bibfield  {author} {\bibinfo {author} {\bibfnamefont {S.}~\bibnamefont
  {Horiuchi}}, \bibinfo {author} {\bibfnamefont {P.~J.}\ \bibnamefont
  {Humphrey}}, \bibinfo {author} {\bibfnamefont {J.}~\bibnamefont {Onorbe}},
  \bibinfo {author} {\bibfnamefont {K.~N.}\ \bibnamefont {Abazajian}}, \bibinfo
  {author} {\bibfnamefont {M.}~\bibnamefont {Kaplinghat}}, \ and\ \bibinfo
  {author} {\bibfnamefont {S.}~\bibnamefont {Garrison-Kimmel}},\ }\href
  {\doibase 10.1103/PhysRevD.89.025017} {\bibfield  {journal} {\bibinfo
  {journal} {Phys. Rev. D}\ }\textbf {\bibinfo {volume} {89}},\ \bibinfo
  {pages} {025017} (\bibinfo {year} {2014})},\ \Eprint
  {http://arxiv.org/abs/1311.0282} {arXiv:1311.0282 [astro-ph.CO]} \BibitemShut
  {NoStop}%
\bibitem [{\citenamefont {Kennedy}\ \emph {et~al.}(2014)\citenamefont
  {Kennedy}, \citenamefont {Frenk}, \citenamefont {Cole},\ and\ \citenamefont
  {Benson}}]{Kennedy:2013uta}%
  \BibitemOpen
  \bibfield  {author} {\bibinfo {author} {\bibfnamefont {R.}~\bibnamefont
  {Kennedy}}, \bibinfo {author} {\bibfnamefont {C.}~\bibnamefont {Frenk}},
  \bibinfo {author} {\bibfnamefont {S.}~\bibnamefont {Cole}}, \ and\ \bibinfo
  {author} {\bibfnamefont {A.}~\bibnamefont {Benson}},\ }\href {\doibase
  10.1093/mnras/stu719} {\bibfield  {journal} {\bibinfo  {journal} {Mon. Not.
  Roy. Astron. Soc.}\ }\textbf {\bibinfo {volume} {442}},\ \bibinfo {pages}
  {2487} (\bibinfo {year} {2014})},\ \Eprint {http://arxiv.org/abs/1310.7739}
  {arXiv:1310.7739 [astro-ph.CO]} \BibitemShut {NoStop}%
\bibitem [{\citenamefont {Schneider}(2015)}]{Schneider:2014rda}%
  \BibitemOpen
  \bibfield  {author} {\bibinfo {author} {\bibfnamefont {A.}~\bibnamefont
  {Schneider}},\ }\href {\doibase 10.1093/mnras/stv1169} {\bibfield  {journal}
  {\bibinfo  {journal} {Mon. Not. Roy. Astron. Soc.}\ }\textbf {\bibinfo
  {volume} {451}},\ \bibinfo {pages} {3117} (\bibinfo {year} {2015})},\ \Eprint
  {http://arxiv.org/abs/1412.2133} {arXiv:1412.2133 [astro-ph.CO]} \BibitemShut
  {NoStop}%
\bibitem [{\citenamefont {Schneider}(2016)}]{Schneider:2016uqi}%
  \BibitemOpen
  \bibfield  {author} {\bibinfo {author} {\bibfnamefont {A.}~\bibnamefont
  {Schneider}},\ }\href {\doibase 10.1088/1475-7516/2016/04/059} {\bibfield
  {journal} {\bibinfo  {journal} {JCAP}\ }\textbf {\bibinfo {volume} {04}},\
  \bibinfo {pages} {059} (\bibinfo {year} {2016})},\ \Eprint
  {http://arxiv.org/abs/1601.07553} {arXiv:1601.07553 [astro-ph.CO]}
  \BibitemShut {NoStop}%
\bibitem [{\citenamefont {Gariazzo}\ \emph {et~al.}(2017)\citenamefont
  {Gariazzo}, \citenamefont {Escudero}, \citenamefont {Diamanti},\ and\
  \citenamefont {Mena}}]{Gariazzo:2017pzb}%
  \BibitemOpen
  \bibfield  {author} {\bibinfo {author} {\bibfnamefont {S.}~\bibnamefont
  {Gariazzo}}, \bibinfo {author} {\bibfnamefont {M.}~\bibnamefont {Escudero}},
  \bibinfo {author} {\bibfnamefont {R.}~\bibnamefont {Diamanti}}, \ and\
  \bibinfo {author} {\bibfnamefont {O.}~\bibnamefont {Mena}},\ }\href {\doibase
  10.1103/PhysRevD.96.043501} {\bibfield  {journal} {\bibinfo  {journal} {Phys.
  Rev. D}\ }\textbf {\bibinfo {volume} {96}},\ \bibinfo {pages} {043501}
  (\bibinfo {year} {2017})},\ \Eprint {http://arxiv.org/abs/1704.02991}
  {arXiv:1704.02991 [astro-ph.CO]} \BibitemShut {NoStop}%
\bibitem [{\citenamefont {Diamanti}\ \emph {et~al.}(2017)\citenamefont
  {Diamanti}, \citenamefont {Ando}, \citenamefont {Gariazzo}, \citenamefont
  {Mena},\ and\ \citenamefont {Weniger}}]{Diamanti:2017xfo}%
  \BibitemOpen
  \bibfield  {author} {\bibinfo {author} {\bibfnamefont {R.}~\bibnamefont
  {Diamanti}}, \bibinfo {author} {\bibfnamefont {S.}~\bibnamefont {Ando}},
  \bibinfo {author} {\bibfnamefont {S.}~\bibnamefont {Gariazzo}}, \bibinfo
  {author} {\bibfnamefont {O.}~\bibnamefont {Mena}}, \ and\ \bibinfo {author}
  {\bibfnamefont {C.}~\bibnamefont {Weniger}},\ }\href {\doibase
  10.1088/1475-7516/2017/06/008} {\bibfield  {journal} {\bibinfo  {journal}
  {JCAP}\ }\textbf {\bibinfo {volume} {06}},\ \bibinfo {pages} {008} (\bibinfo
  {year} {2017})},\ \Eprint {http://arxiv.org/abs/1701.03128} {arXiv:1701.03128
  [astro-ph.CO]} \BibitemShut {NoStop}%
\bibitem [{\citenamefont {D'Eramo}\ and\ \citenamefont
  {Lenoci}(2020)}]{DEramo:2020gpr}%
  \BibitemOpen
  \bibfield  {author} {\bibinfo {author} {\bibfnamefont {F.}~\bibnamefont
  {D'Eramo}}\ and\ \bibinfo {author} {\bibfnamefont {A.}~\bibnamefont
  {Lenoci}},\ }\href@noop {} {\  (\bibinfo {year} {2020})},\ \Eprint
  {http://arxiv.org/abs/2012.01446} {arXiv:2012.01446 [hep-ph]} \BibitemShut
  {NoStop}%
\bibitem [{\citenamefont {Press}\ and\ \citenamefont
  {Schechter}(1974)}]{Press:1973iz}%
  \BibitemOpen
  \bibfield  {author} {\bibinfo {author} {\bibfnamefont {W.~H.}\ \bibnamefont
  {Press}}\ and\ \bibinfo {author} {\bibfnamefont {P.}~\bibnamefont
  {Schechter}},\ }\href {\doibase 10.1086/152650} {\bibfield  {journal}
  {\bibinfo  {journal} {Astrophys. J.}\ }\textbf {\bibinfo {volume} {187}},\
  \bibinfo {pages} {425} (\bibinfo {year} {1974})}\BibitemShut {NoStop}%
\bibitem [{\citenamefont {Bond}\ \emph {et~al.}(1991)\citenamefont {Bond},
  \citenamefont {Cole}, \citenamefont {Efstathiou},\ and\ \citenamefont
  {Kaiser}}]{Bond:1990iw}%
  \BibitemOpen
  \bibfield  {author} {\bibinfo {author} {\bibfnamefont {J.~R.}\ \bibnamefont
  {Bond}}, \bibinfo {author} {\bibfnamefont {S.}~\bibnamefont {Cole}}, \bibinfo
  {author} {\bibfnamefont {G.}~\bibnamefont {Efstathiou}}, \ and\ \bibinfo
  {author} {\bibfnamefont {N.}~\bibnamefont {Kaiser}},\ }\href {\doibase
  10.1086/170520} {\bibfield  {journal} {\bibinfo  {journal} {Astrophys. J.}\
  }\textbf {\bibinfo {volume} {379}},\ \bibinfo {pages} {440} (\bibinfo {year}
  {1991})}\BibitemShut {NoStop}%
\bibitem [{\citenamefont {Lacey}\ and\ \citenamefont
  {Cole}(1993)}]{Lacey:1993iv}%
  \BibitemOpen
  \bibfield  {author} {\bibinfo {author} {\bibfnamefont {C.~G.}\ \bibnamefont
  {Lacey}}\ and\ \bibinfo {author} {\bibfnamefont {S.}~\bibnamefont {Cole}},\
  }\href@noop {} {\bibfield  {journal} {\bibinfo  {journal} {Mon. Not. Roy.
  Astron. Soc.}\ }\textbf {\bibinfo {volume} {262}},\ \bibinfo {pages} {627}
  (\bibinfo {year} {1993})}\BibitemShut {NoStop}%
\bibitem [{\citenamefont {Lovell}\ \emph {et~al.}(2014)\citenamefont {Lovell},
  \citenamefont {Frenk}, \citenamefont {Eke}, \citenamefont {Jenkins},
  \citenamefont {Gao},\ and\ \citenamefont {Theuns}}]{Lovell:2013ola}%
  \BibitemOpen
  \bibfield  {author} {\bibinfo {author} {\bibfnamefont {M.~R.}\ \bibnamefont
  {Lovell}}, \bibinfo {author} {\bibfnamefont {C.~S.}\ \bibnamefont {Frenk}},
  \bibinfo {author} {\bibfnamefont {V.~R.}\ \bibnamefont {Eke}}, \bibinfo
  {author} {\bibfnamefont {A.}~\bibnamefont {Jenkins}}, \bibinfo {author}
  {\bibfnamefont {L.}~\bibnamefont {Gao}}, \ and\ \bibinfo {author}
  {\bibfnamefont {T.}~\bibnamefont {Theuns}},\ }\href {\doibase
  10.1093/mnras/stt2431} {\bibfield  {journal} {\bibinfo  {journal} {Mon. Not.
  Roy. Astron. Soc.}\ }\textbf {\bibinfo {volume} {439}},\ \bibinfo {pages}
  {300} (\bibinfo {year} {2014})},\ \Eprint {http://arxiv.org/abs/1308.1399}
  {arXiv:1308.1399 [astro-ph.CO]} \BibitemShut {NoStop}%
\bibitem [{\citenamefont {Brooks}\ and\ \citenamefont
  {Zolotov}(2014)}]{Brooks:2012vi}%
  \BibitemOpen
  \bibfield  {author} {\bibinfo {author} {\bibfnamefont {A.~M.}\ \bibnamefont
  {Brooks}}\ and\ \bibinfo {author} {\bibfnamefont {A.}~\bibnamefont
  {Zolotov}},\ }\href {\doibase 10.1088/0004-637X/786/2/87} {\bibfield
  {journal} {\bibinfo  {journal} {Astrophys. J.}\ }\textbf {\bibinfo {volume}
  {786}},\ \bibinfo {pages} {87} (\bibinfo {year} {2014})},\ \Eprint
  {http://arxiv.org/abs/1207.2468} {arXiv:1207.2468 [astro-ph.CO]} \BibitemShut
  {NoStop}%
\bibitem [{\citenamefont {Guo}\ \emph {et~al.}(2010)\citenamefont {Guo},
  \citenamefont {White}, \citenamefont {Li},\ and\ \citenamefont
  {Boylan-Kolchin}}]{Guo:2009fn}%
  \BibitemOpen
  \bibfield  {author} {\bibinfo {author} {\bibfnamefont {Q.}~\bibnamefont
  {Guo}}, \bibinfo {author} {\bibfnamefont {S.}~\bibnamefont {White}}, \bibinfo
  {author} {\bibfnamefont {C.}~\bibnamefont {Li}}, \ and\ \bibinfo {author}
  {\bibfnamefont {M.}~\bibnamefont {Boylan-Kolchin}},\ }\href {\doibase
  10.1111/j.1365-2966.2010.16341.x} {\bibfield  {journal} {\bibinfo  {journal}
  {Mon. Not. Roy. Astron. Soc.}\ }\textbf {\bibinfo {volume} {404}},\ \bibinfo
  {pages} {1111} (\bibinfo {year} {2010})},\ \Eprint
  {http://arxiv.org/abs/0909.4305} {arXiv:0909.4305 [astro-ph.CO]} \BibitemShut
  {NoStop}%
\bibitem [{\citenamefont {Murgia}\ \emph {et~al.}(2018)\citenamefont {Murgia},
  \citenamefont {Ir\v{s}i\v{c}},\ and\ \citenamefont {Viel}}]{Murgia:2018now}%
  \BibitemOpen
  \bibfield  {author} {\bibinfo {author} {\bibfnamefont {R.}~\bibnamefont
  {Murgia}}, \bibinfo {author} {\bibfnamefont {V.}~\bibnamefont
  {Ir\v{s}i\v{c}}}, \ and\ \bibinfo {author} {\bibfnamefont {M.}~\bibnamefont
  {Viel}},\ }\href {\doibase 10.1103/PhysRevD.98.083540} {\bibfield  {journal}
  {\bibinfo  {journal} {Phys. Rev. D}\ }\textbf {\bibinfo {volume} {98}},\
  \bibinfo {pages} {083540} (\bibinfo {year} {2018})},\ \Eprint
  {http://arxiv.org/abs/1806.08371} {arXiv:1806.08371 [astro-ph.CO]}
  \BibitemShut {NoStop}%
\bibitem [{\citenamefont {Baur}\ \emph {et~al.}(2017)\citenamefont {Baur},
  \citenamefont {Palanque-Delabrouille}, \citenamefont {Yeche}, \citenamefont
  {Boyarsky}, \citenamefont {Ruchayskiy}, \citenamefont {Armengaud},\ and\
  \citenamefont {Lesgourgues}}]{Baur:2017stq}%
  \BibitemOpen
  \bibfield  {author} {\bibinfo {author} {\bibfnamefont {J.}~\bibnamefont
  {Baur}}, \bibinfo {author} {\bibfnamefont {N.}~\bibnamefont
  {Palanque-Delabrouille}}, \bibinfo {author} {\bibfnamefont {C.}~\bibnamefont
  {Yeche}}, \bibinfo {author} {\bibfnamefont {A.}~\bibnamefont {Boyarsky}},
  \bibinfo {author} {\bibfnamefont {O.}~\bibnamefont {Ruchayskiy}}, \bibinfo
  {author} {\bibfnamefont {E.}~\bibnamefont {Armengaud}}, \ and\ \bibinfo
  {author} {\bibfnamefont {J.}~\bibnamefont {Lesgourgues}},\ }\href {\doibase
  10.1088/1475-7516/2017/12/013} {\bibfield  {journal} {\bibinfo  {journal}
  {JCAP}\ }\textbf {\bibinfo {volume} {12}},\ \bibinfo {pages} {013} (\bibinfo
  {year} {2017})},\ \Eprint {http://arxiv.org/abs/1706.03118} {arXiv:1706.03118
  [astro-ph.CO]} \BibitemShut {NoStop}%
\bibitem [{\citenamefont {Akrami}\ \emph {et~al.}(2020)\citenamefont {Akrami}
  \emph {et~al.}}]{Akrami:2018odb}%
  \BibitemOpen
  \bibfield  {author} {\bibinfo {author} {\bibfnamefont {Y.}~\bibnamefont
  {Akrami}} \emph {et~al.} (\bibinfo {collaboration} {Planck}),\ }\href
  {\doibase 10.1051/0004-6361/201833887} {\bibfield  {journal} {\bibinfo
  {journal} {Astron. Astrophys.}\ }\textbf {\bibinfo {volume} {641}},\ \bibinfo
  {pages} {A10} (\bibinfo {year} {2020})},\ \Eprint
  {http://arxiv.org/abs/1807.06211} {arXiv:1807.06211 [astro-ph.CO]}
  \BibitemShut {NoStop}%
\bibitem [{\citenamefont {Schwarz}(1998)}]{Schwarz:1997gv}%
  \BibitemOpen
  \bibfield  {author} {\bibinfo {author} {\bibfnamefont {D.~J.}\ \bibnamefont
  {Schwarz}},\ }\href {\doibase 10.1142/S0217732398002941} {\bibfield
  {journal} {\bibinfo  {journal} {Mod. Phys. Lett. A}\ }\textbf {\bibinfo
  {volume} {13}},\ \bibinfo {pages} {2771} (\bibinfo {year} {1998})},\ \Eprint
  {http://arxiv.org/abs/gr-qc/9709027} {arXiv:gr-qc/9709027} \BibitemShut
  {NoStop}%
\bibitem [{\citenamefont {Watanabe}\ and\ \citenamefont
  {Komatsu}(2006)}]{Watanabe:2006qe}%
  \BibitemOpen
  \bibfield  {author} {\bibinfo {author} {\bibfnamefont {Y.}~\bibnamefont
  {Watanabe}}\ and\ \bibinfo {author} {\bibfnamefont {E.}~\bibnamefont
  {Komatsu}},\ }\href {\doibase 10.1103/PhysRevD.73.123515} {\bibfield
  {journal} {\bibinfo  {journal} {Phys. Rev. D}\ }\textbf {\bibinfo {volume}
  {73}},\ \bibinfo {pages} {123515} (\bibinfo {year} {2006})},\ \Eprint
  {http://arxiv.org/abs/astro-ph/0604176} {arXiv:astro-ph/0604176} \BibitemShut
  {NoStop}%
\bibitem [{\citenamefont {Kuroyanagi}\ \emph {et~al.}(2009)\citenamefont
  {Kuroyanagi}, \citenamefont {Chiba},\ and\ \citenamefont
  {Sugiyama}}]{Kuroyanagi:2008ye}%
  \BibitemOpen
  \bibfield  {author} {\bibinfo {author} {\bibfnamefont {S.}~\bibnamefont
  {Kuroyanagi}}, \bibinfo {author} {\bibfnamefont {T.}~\bibnamefont {Chiba}}, \
  and\ \bibinfo {author} {\bibfnamefont {N.}~\bibnamefont {Sugiyama}},\ }\href
  {\doibase 10.1103/PhysRevD.79.103501} {\bibfield  {journal} {\bibinfo
  {journal} {Phys. Rev. D}\ }\textbf {\bibinfo {volume} {79}},\ \bibinfo
  {pages} {103501} (\bibinfo {year} {2009})},\ \Eprint
  {http://arxiv.org/abs/0804.3249} {arXiv:0804.3249 [astro-ph]} \BibitemShut
  {NoStop}%
\bibitem [{\citenamefont {Saikawa}\ and\ \citenamefont
  {Shirai}(2018)}]{Saikawa:2018rcs}%
  \BibitemOpen
  \bibfield  {author} {\bibinfo {author} {\bibfnamefont {K.}~\bibnamefont
  {Saikawa}}\ and\ \bibinfo {author} {\bibfnamefont {S.}~\bibnamefont
  {Shirai}},\ }\href {\doibase 10.1088/1475-7516/2018/05/035} {\bibfield
  {journal} {\bibinfo  {journal} {JCAP}\ }\textbf {\bibinfo {volume} {05}},\
  \bibinfo {pages} {035} (\bibinfo {year} {2018})},\ \Eprint
  {http://arxiv.org/abs/1803.01038} {arXiv:1803.01038 [hep-ph]} \BibitemShut
  {NoStop}%
\bibitem [{\citenamefont {Seto}\ and\ \citenamefont
  {Yokoyama}(2003)}]{Seto:2003kc}%
  \BibitemOpen
  \bibfield  {author} {\bibinfo {author} {\bibfnamefont {N.}~\bibnamefont
  {Seto}}\ and\ \bibinfo {author} {\bibfnamefont {J.}~\bibnamefont
  {Yokoyama}},\ }\href {\doibase 10.1143/JPSJ.72.3082} {\bibfield  {journal}
  {\bibinfo  {journal} {J. Phys. Soc. Jap.}\ }\textbf {\bibinfo {volume}
  {72}},\ \bibinfo {pages} {3082} (\bibinfo {year} {2003})},\ \Eprint
  {http://arxiv.org/abs/gr-qc/0305096} {arXiv:gr-qc/0305096} \BibitemShut
  {NoStop}%
\bibitem [{\citenamefont {Boyle}\ and\ \citenamefont
  {Steinhardt}(2008)}]{Boyle:2005se}%
  \BibitemOpen
  \bibfield  {author} {\bibinfo {author} {\bibfnamefont {L.~A.}\ \bibnamefont
  {Boyle}}\ and\ \bibinfo {author} {\bibfnamefont {P.~J.}\ \bibnamefont
  {Steinhardt}},\ }\href {\doibase 10.1103/PhysRevD.77.063504} {\bibfield
  {journal} {\bibinfo  {journal} {Phys. Rev. D}\ }\textbf {\bibinfo {volume}
  {77}},\ \bibinfo {pages} {063504} (\bibinfo {year} {2008})},\ \Eprint
  {http://arxiv.org/abs/astro-ph/0512014} {arXiv:astro-ph/0512014} \BibitemShut
  {NoStop}%
\bibitem [{\citenamefont {Jinno}\ \emph
  {et~al.}(2012{\natexlab{a}})\citenamefont {Jinno}, \citenamefont {Moroi},\
  and\ \citenamefont {Nakayama}}]{Jinno:2011sw}%
  \BibitemOpen
  \bibfield  {author} {\bibinfo {author} {\bibfnamefont {R.}~\bibnamefont
  {Jinno}}, \bibinfo {author} {\bibfnamefont {T.}~\bibnamefont {Moroi}}, \ and\
  \bibinfo {author} {\bibfnamefont {K.}~\bibnamefont {Nakayama}},\ }\href
  {\doibase 10.1016/j.physletb.2012.05.061} {\bibfield  {journal} {\bibinfo
  {journal} {Phys. Lett. B}\ }\textbf {\bibinfo {volume} {713}},\ \bibinfo
  {pages} {129} (\bibinfo {year} {2012}{\natexlab{a}})},\ \Eprint
  {http://arxiv.org/abs/1112.0084} {arXiv:1112.0084 [hep-ph]} \BibitemShut
  {NoStop}%
\bibitem [{\citenamefont {Jinno}\ \emph
  {et~al.}(2012{\natexlab{b}})\citenamefont {Jinno}, \citenamefont {Moroi},\
  and\ \citenamefont {Nakayama}}]{Jinno:2012xb}%
  \BibitemOpen
  \bibfield  {author} {\bibinfo {author} {\bibfnamefont {R.}~\bibnamefont
  {Jinno}}, \bibinfo {author} {\bibfnamefont {T.}~\bibnamefont {Moroi}}, \ and\
  \bibinfo {author} {\bibfnamefont {K.}~\bibnamefont {Nakayama}},\ }\href
  {\doibase 10.1103/PhysRevD.86.123502} {\bibfield  {journal} {\bibinfo
  {journal} {Phys. Rev. D}\ }\textbf {\bibinfo {volume} {86}},\ \bibinfo
  {pages} {123502} (\bibinfo {year} {2012}{\natexlab{b}})},\ \Eprint
  {http://arxiv.org/abs/1208.0184} {arXiv:1208.0184 [astro-ph.CO]} \BibitemShut
  {NoStop}%
\bibitem [{\citenamefont {Jinno}\ \emph
  {et~al.}(2014{\natexlab{a}})\citenamefont {Jinno}, \citenamefont {Moroi},\
  and\ \citenamefont {Nakayama}}]{Jinno:2013xqa}%
  \BibitemOpen
  \bibfield  {author} {\bibinfo {author} {\bibfnamefont {R.}~\bibnamefont
  {Jinno}}, \bibinfo {author} {\bibfnamefont {T.}~\bibnamefont {Moroi}}, \ and\
  \bibinfo {author} {\bibfnamefont {K.}~\bibnamefont {Nakayama}},\ }\href
  {\doibase 10.1088/1475-7516/2014/01/040} {\bibfield  {journal} {\bibinfo
  {journal} {JCAP}\ }\textbf {\bibinfo {volume} {01}},\ \bibinfo {pages} {040}
  (\bibinfo {year} {2014}{\natexlab{a}})},\ \Eprint
  {http://arxiv.org/abs/1307.3010} {arXiv:1307.3010 [hep-ph]} \BibitemShut
  {NoStop}%
\bibitem [{\citenamefont {Caldwell}\ \emph {et~al.}(2019)\citenamefont
  {Caldwell}, \citenamefont {Smith},\ and\ \citenamefont
  {Walker}}]{Caldwell:2018giq}%
  \BibitemOpen
  \bibfield  {author} {\bibinfo {author} {\bibfnamefont {R.~R.}\ \bibnamefont
  {Caldwell}}, \bibinfo {author} {\bibfnamefont {T.~L.}\ \bibnamefont {Smith}},
  \ and\ \bibinfo {author} {\bibfnamefont {D.~G.~E.}\ \bibnamefont {Walker}},\
  }\href {\doibase 10.1103/PhysRevD.100.043513} {\bibfield  {journal} {\bibinfo
   {journal} {Phys. Rev. D}\ }\textbf {\bibinfo {volume} {100}},\ \bibinfo
  {pages} {043513} (\bibinfo {year} {2019})},\ \Eprint
  {http://arxiv.org/abs/1812.07577} {arXiv:1812.07577 [astro-ph.CO]}
  \BibitemShut {NoStop}%
\bibitem [{\citenamefont {Ringwald}\ \emph
  {et~al.}(2021{\natexlab{a}})\citenamefont {Ringwald}, \citenamefont
  {Sch\"utte-Engel},\ and\ \citenamefont {Tamarit}}]{Ringwald:2020ist}%
  \BibitemOpen
  \bibfield  {author} {\bibinfo {author} {\bibfnamefont {A.}~\bibnamefont
  {Ringwald}}, \bibinfo {author} {\bibfnamefont {J.}~\bibnamefont
  {Sch\"utte-Engel}}, \ and\ \bibinfo {author} {\bibfnamefont {C.}~\bibnamefont
  {Tamarit}},\ }\href {\doibase 10.1088/1475-7516/2021/03/054} {\bibfield
  {journal} {\bibinfo  {journal} {JCAP}\ }\textbf {\bibinfo {volume} {03}},\
  \bibinfo {pages} {054} (\bibinfo {year} {2021}{\natexlab{a}})},\ \Eprint
  {http://arxiv.org/abs/2011.04731} {arXiv:2011.04731 [hep-ph]} \BibitemShut
  {NoStop}%
\bibitem [{\citenamefont {Ringwald}\ \emph
  {et~al.}(2021{\natexlab{b}})\citenamefont {Ringwald}, \citenamefont
  {Saikawa},\ and\ \citenamefont {Tamarit}}]{Ringwald:2020vei}%
  \BibitemOpen
  \bibfield  {author} {\bibinfo {author} {\bibfnamefont {A.}~\bibnamefont
  {Ringwald}}, \bibinfo {author} {\bibfnamefont {K.}~\bibnamefont {Saikawa}}, \
  and\ \bibinfo {author} {\bibfnamefont {C.}~\bibnamefont {Tamarit}},\ }\href
  {\doibase 10.1088/1475-7516/2021/02/046} {\bibfield  {journal} {\bibinfo
  {journal} {JCAP}\ }\textbf {\bibinfo {volume} {02}},\ \bibinfo {pages} {046}
  (\bibinfo {year} {2021}{\natexlab{b}})},\ \Eprint
  {http://arxiv.org/abs/2009.02050} {arXiv:2009.02050 [hep-ph]} \BibitemShut
  {NoStop}%
\bibitem [{\citenamefont {Nakayama}\ \emph {et~al.}(2008)\citenamefont
  {Nakayama}, \citenamefont {Saito}, \citenamefont {Suwa},\ and\ \citenamefont
  {Yokoyama}}]{Nakayama:2008ip}%
  \BibitemOpen
  \bibfield  {author} {\bibinfo {author} {\bibfnamefont {K.}~\bibnamefont
  {Nakayama}}, \bibinfo {author} {\bibfnamefont {S.}~\bibnamefont {Saito}},
  \bibinfo {author} {\bibfnamefont {Y.}~\bibnamefont {Suwa}}, \ and\ \bibinfo
  {author} {\bibfnamefont {J.}~\bibnamefont {Yokoyama}},\ }\href {\doibase
  10.1103/PhysRevD.77.124001} {\bibfield  {journal} {\bibinfo  {journal} {Phys.
  Rev. D}\ }\textbf {\bibinfo {volume} {77}},\ \bibinfo {pages} {124001}
  (\bibinfo {year} {2008})},\ \Eprint {http://arxiv.org/abs/0802.2452}
  {arXiv:0802.2452 [hep-ph]} \BibitemShut {NoStop}%
\bibitem [{\citenamefont {Kuroyanagi}\ \emph {et~al.}(2011)\citenamefont
  {Kuroyanagi}, \citenamefont {Nakayama},\ and\ \citenamefont
  {Saito}}]{Kuroyanagi:2011fy}%
  \BibitemOpen
  \bibfield  {author} {\bibinfo {author} {\bibfnamefont {S.}~\bibnamefont
  {Kuroyanagi}}, \bibinfo {author} {\bibfnamefont {K.}~\bibnamefont
  {Nakayama}}, \ and\ \bibinfo {author} {\bibfnamefont {S.}~\bibnamefont
  {Saito}},\ }\href {\doibase 10.1103/PhysRevD.84.123513} {\bibfield  {journal}
  {\bibinfo  {journal} {Phys. Rev. D}\ }\textbf {\bibinfo {volume} {84}},\
  \bibinfo {pages} {123513} (\bibinfo {year} {2011})},\ \Eprint
  {http://arxiv.org/abs/1110.4169} {arXiv:1110.4169 [astro-ph.CO]} \BibitemShut
  {NoStop}%
\bibitem [{\citenamefont {Jinno}\ \emph
  {et~al.}(2014{\natexlab{b}})\citenamefont {Jinno}, \citenamefont {Moroi},\
  and\ \citenamefont {Takahashi}}]{Jinno:2014qka}%
  \BibitemOpen
  \bibfield  {author} {\bibinfo {author} {\bibfnamefont {R.}~\bibnamefont
  {Jinno}}, \bibinfo {author} {\bibfnamefont {T.}~\bibnamefont {Moroi}}, \ and\
  \bibinfo {author} {\bibfnamefont {T.}~\bibnamefont {Takahashi}},\ }\href
  {\doibase 10.1088/1475-7516/2014/12/006} {\bibfield  {journal} {\bibinfo
  {journal} {JCAP}\ }\textbf {\bibinfo {volume} {12}},\ \bibinfo {pages} {006}
  (\bibinfo {year} {2014}{\natexlab{b}})},\ \Eprint
  {http://arxiv.org/abs/1406.1666} {arXiv:1406.1666 [astro-ph.CO]} \BibitemShut
  {NoStop}%
\bibitem [{\citenamefont {Kuroyanagi}\ \emph {et~al.}(2015)\citenamefont
  {Kuroyanagi}, \citenamefont {Takahashi},\ and\ \citenamefont
  {Yokoyama}}]{Kuroyanagi:2014nba}%
  \BibitemOpen
  \bibfield  {author} {\bibinfo {author} {\bibfnamefont {S.}~\bibnamefont
  {Kuroyanagi}}, \bibinfo {author} {\bibfnamefont {T.}~\bibnamefont
  {Takahashi}}, \ and\ \bibinfo {author} {\bibfnamefont {S.}~\bibnamefont
  {Yokoyama}},\ }\href {\doibase 10.1088/1475-7516/2015/02/003} {\bibfield
  {journal} {\bibinfo  {journal} {JCAP}\ }\textbf {\bibinfo {volume} {02}},\
  \bibinfo {pages} {003} (\bibinfo {year} {2015})},\ \Eprint
  {http://arxiv.org/abs/1407.4785} {arXiv:1407.4785 [astro-ph.CO]} \BibitemShut
  {NoStop}%
\bibitem [{\citenamefont {D'Eramo}\ and\ \citenamefont
  {Schmitz}(2019)}]{DEramo:2019tit}%
  \BibitemOpen
  \bibfield  {author} {\bibinfo {author} {\bibfnamefont {F.}~\bibnamefont
  {D'Eramo}}\ and\ \bibinfo {author} {\bibfnamefont {K.}~\bibnamefont
  {Schmitz}},\ }\href {\doibase 10.1103/PhysRevResearch.1.013010} {\bibfield
  {journal} {\bibinfo  {journal} {Phys. Rev. Research.}\ }\textbf {\bibinfo
  {volume} {1}},\ \bibinfo {pages} {013010} (\bibinfo {year} {2019})},\ \Eprint
  {http://arxiv.org/abs/1904.07870} {arXiv:1904.07870 [hep-ph]} \BibitemShut
  {NoStop}%
\bibitem [{\citenamefont {Cook}\ and\ \citenamefont
  {Sorbo}(2012)}]{Cook:2011hg}%
  \BibitemOpen
  \bibfield  {author} {\bibinfo {author} {\bibfnamefont {J.~L.}\ \bibnamefont
  {Cook}}\ and\ \bibinfo {author} {\bibfnamefont {L.}~\bibnamefont {Sorbo}},\
  }\href {\doibase 10.1103/PhysRevD.85.023534} {\bibfield  {journal} {\bibinfo
  {journal} {Phys. Rev. D}\ }\textbf {\bibinfo {volume} {85}},\ \bibinfo
  {pages} {023534} (\bibinfo {year} {2012})},\ \bibinfo {note} {[Erratum:
  Phys.Rev.D 86, 069901 (2012)]},\ \Eprint {http://arxiv.org/abs/1109.0022}
  {arXiv:1109.0022 [astro-ph.CO]} \BibitemShut {NoStop}%
\bibitem [{\citenamefont {Barnaby}\ \emph {et~al.}(2012)\citenamefont
  {Barnaby}, \citenamefont {Pajer},\ and\ \citenamefont
  {Peloso}}]{Barnaby:2011qe}%
  \BibitemOpen
  \bibfield  {author} {\bibinfo {author} {\bibfnamefont {N.}~\bibnamefont
  {Barnaby}}, \bibinfo {author} {\bibfnamefont {E.}~\bibnamefont {Pajer}}, \
  and\ \bibinfo {author} {\bibfnamefont {M.}~\bibnamefont {Peloso}},\ }\href
  {\doibase 10.1103/PhysRevD.85.023525} {\bibfield  {journal} {\bibinfo
  {journal} {Phys. Rev. D}\ }\textbf {\bibinfo {volume} {85}},\ \bibinfo
  {pages} {023525} (\bibinfo {year} {2012})},\ \Eprint
  {http://arxiv.org/abs/1110.3327} {arXiv:1110.3327 [astro-ph.CO]} \BibitemShut
  {NoStop}%
\bibitem [{\citenamefont {Anber}\ and\ \citenamefont
  {Sorbo}(2012)}]{Anber:2012du}%
  \BibitemOpen
  \bibfield  {author} {\bibinfo {author} {\bibfnamefont {M.~M.}\ \bibnamefont
  {Anber}}\ and\ \bibinfo {author} {\bibfnamefont {L.}~\bibnamefont {Sorbo}},\
  }\href {\doibase 10.1103/PhysRevD.85.123537} {\bibfield  {journal} {\bibinfo
  {journal} {Phys. Rev. D}\ }\textbf {\bibinfo {volume} {85}},\ \bibinfo
  {pages} {123537} (\bibinfo {year} {2012})},\ \Eprint
  {http://arxiv.org/abs/1203.5849} {arXiv:1203.5849 [astro-ph.CO]} \BibitemShut
  {NoStop}%
\bibitem [{\citenamefont {Domcke}\ \emph {et~al.}(2016)\citenamefont {Domcke},
  \citenamefont {Pieroni},\ and\ \citenamefont {Bin\'etruy}}]{Domcke:2016bkh}%
  \BibitemOpen
  \bibfield  {author} {\bibinfo {author} {\bibfnamefont {V.}~\bibnamefont
  {Domcke}}, \bibinfo {author} {\bibfnamefont {M.}~\bibnamefont {Pieroni}}, \
  and\ \bibinfo {author} {\bibfnamefont {P.}~\bibnamefont {Bin\'etruy}},\
  }\href {\doibase 10.1088/1475-7516/2016/06/031} {\bibfield  {journal}
  {\bibinfo  {journal} {JCAP}\ }\textbf {\bibinfo {volume} {06}},\ \bibinfo
  {pages} {031} (\bibinfo {year} {2016})},\ \Eprint
  {http://arxiv.org/abs/1603.01287} {arXiv:1603.01287 [astro-ph.CO]}
  \BibitemShut {NoStop}%
\bibitem [{\citenamefont {Jim\'enez}\ \emph {et~al.}(2017)\citenamefont
  {Jim\'enez}, \citenamefont {Kamada}, \citenamefont {Schmitz},\ and\
  \citenamefont {Xu}}]{Jimenez:2017cdr}%
  \BibitemOpen
  \bibfield  {author} {\bibinfo {author} {\bibfnamefont {D.}~\bibnamefont
  {Jim\'enez}}, \bibinfo {author} {\bibfnamefont {K.}~\bibnamefont {Kamada}},
  \bibinfo {author} {\bibfnamefont {K.}~\bibnamefont {Schmitz}}, \ and\
  \bibinfo {author} {\bibfnamefont {X.-J.}\ \bibnamefont {Xu}},\ }\href
  {\doibase 10.1088/1475-7516/2017/12/011} {\bibfield  {journal} {\bibinfo
  {journal} {JCAP}\ }\textbf {\bibinfo {volume} {12}},\ \bibinfo {pages} {011}
  (\bibinfo {year} {2017})},\ \Eprint {http://arxiv.org/abs/1707.07943}
  {arXiv:1707.07943 [hep-ph]} \BibitemShut {NoStop}%
\bibitem [{\citenamefont {Papageorgiou}\ \emph {et~al.}(2019)\citenamefont
  {Papageorgiou}, \citenamefont {Peloso},\ and\ \citenamefont
  {Unal}}]{Papageorgiou:2019ecb}%
  \BibitemOpen
  \bibfield  {author} {\bibinfo {author} {\bibfnamefont {A.}~\bibnamefont
  {Papageorgiou}}, \bibinfo {author} {\bibfnamefont {M.}~\bibnamefont
  {Peloso}}, \ and\ \bibinfo {author} {\bibfnamefont {C.}~\bibnamefont
  {Unal}},\ }\href {\doibase 10.1088/1475-7516/2019/07/004} {\bibfield
  {journal} {\bibinfo  {journal} {JCAP}\ }\textbf {\bibinfo {volume} {07}},\
  \bibinfo {pages} {004} (\bibinfo {year} {2019})},\ \Eprint
  {http://arxiv.org/abs/1904.01488} {arXiv:1904.01488 [astro-ph.CO]}
  \BibitemShut {NoStop}%
\bibitem [{\citenamefont {Abbott}\ \emph {et~al.}(2017)\citenamefont {Abbott}
  \emph {et~al.}}]{TheLIGOScientific:2016dpb}%
  \BibitemOpen
  \bibfield  {author} {\bibinfo {author} {\bibfnamefont {B.~P.}\ \bibnamefont
  {Abbott}} \emph {et~al.} (\bibinfo {collaboration} {LIGO Scientific,
  Virgo}),\ }\href {\doibase 10.1103/PhysRevLett.118.121101} {\bibfield
  {journal} {\bibinfo  {journal} {Phys. Rev. Lett.}\ }\textbf {\bibinfo
  {volume} {118}},\ \bibinfo {pages} {121101} (\bibinfo {year} {2017})},\
  \bibinfo {note} {[Erratum: Phys.Rev.Lett. 119, 029901 (2017)]},\ \Eprint
  {http://arxiv.org/abs/1612.02029} {arXiv:1612.02029 [gr-qc]} \BibitemShut
  {NoStop}%
\bibitem [{\citenamefont {Abbott}\ \emph {et~al.}(2019)\citenamefont {Abbott}
  \emph {et~al.}}]{LIGOScientific:2019vic}%
  \BibitemOpen
  \bibfield  {author} {\bibinfo {author} {\bibfnamefont {B.~P.}\ \bibnamefont
  {Abbott}} \emph {et~al.} (\bibinfo {collaboration} {LIGO Scientific,
  Virgo}),\ }\href {\doibase 10.1103/PhysRevD.100.061101} {\bibfield  {journal}
  {\bibinfo  {journal} {Phys. Rev. D}\ }\textbf {\bibinfo {volume} {100}},\
  \bibinfo {pages} {061101} (\bibinfo {year} {2019})},\ \Eprint
  {http://arxiv.org/abs/1903.02886} {arXiv:1903.02886 [gr-qc]} \BibitemShut
  {NoStop}%
\bibitem [{\citenamefont {Kawasaki}\ \emph {et~al.}(2005)\citenamefont
  {Kawasaki}, \citenamefont {Kohri},\ and\ \citenamefont
  {Moroi}}]{Kawasaki:2004qu}%
  \BibitemOpen
  \bibfield  {author} {\bibinfo {author} {\bibfnamefont {M.}~\bibnamefont
  {Kawasaki}}, \bibinfo {author} {\bibfnamefont {K.}~\bibnamefont {Kohri}}, \
  and\ \bibinfo {author} {\bibfnamefont {T.}~\bibnamefont {Moroi}},\ }\href
  {\doibase 10.1103/PhysRevD.71.083502} {\bibfield  {journal} {\bibinfo
  {journal} {Phys. Rev. D}\ }\textbf {\bibinfo {volume} {71}},\ \bibinfo
  {pages} {083502} (\bibinfo {year} {2005})},\ \Eprint
  {http://arxiv.org/abs/astro-ph/0408426} {arXiv:astro-ph/0408426} \BibitemShut
  {NoStop}%
\bibitem [{\citenamefont {Hasegawa}\ \emph {et~al.}(2019)\citenamefont
  {Hasegawa}, \citenamefont {Hiroshima}, \citenamefont {Kohri}, \citenamefont
  {Hansen}, \citenamefont {Tram},\ and\ \citenamefont
  {Hannestad}}]{Hasegawa:2019jsa}%
  \BibitemOpen
  \bibfield  {author} {\bibinfo {author} {\bibfnamefont {T.}~\bibnamefont
  {Hasegawa}}, \bibinfo {author} {\bibfnamefont {N.}~\bibnamefont {Hiroshima}},
  \bibinfo {author} {\bibfnamefont {K.}~\bibnamefont {Kohri}}, \bibinfo
  {author} {\bibfnamefont {R.~S.~L.}\ \bibnamefont {Hansen}}, \bibinfo {author}
  {\bibfnamefont {T.}~\bibnamefont {Tram}}, \ and\ \bibinfo {author}
  {\bibfnamefont {S.}~\bibnamefont {Hannestad}},\ }\href {\doibase
  10.1088/1475-7516/2019/12/012} {\bibfield  {journal} {\bibinfo  {journal}
  {JCAP}\ }\textbf {\bibinfo {volume} {12}},\ \bibinfo {pages} {012} (\bibinfo
  {year} {2019})},\ \Eprint {http://arxiv.org/abs/1908.10189} {arXiv:1908.10189
  [hep-ph]} \BibitemShut {NoStop}%
\bibitem [{\citenamefont {Carilli}\ and\ \citenamefont
  {Rawlings}(2004)}]{Carilli:2004nx}%
  \BibitemOpen
  \bibfield  {author} {\bibinfo {author} {\bibfnamefont {C.~L.}\ \bibnamefont
  {Carilli}}\ and\ \bibinfo {author} {\bibfnamefont {S.}~\bibnamefont
  {Rawlings}},\ }\href {\doibase 10.1016/j.newar.2004.09.001} {\bibfield
  {journal} {\bibinfo  {journal} {New Astron. Rev.}\ }\textbf {\bibinfo
  {volume} {48}},\ \bibinfo {pages} {979} (\bibinfo {year} {2004})},\ \Eprint
  {http://arxiv.org/abs/astro-ph/0409274} {arXiv:astro-ph/0409274} \BibitemShut
  {NoStop}%
\bibitem [{\citenamefont {Janssen}\ \emph {et~al.}(2015)\citenamefont {Janssen}
  \emph {et~al.}}]{Janssen:2014dka}%
  \BibitemOpen
  \bibfield  {author} {\bibinfo {author} {\bibfnamefont {G.}~\bibnamefont
  {Janssen}} \emph {et~al.},\ }\href {\doibase 10.22323/1.215.0037} {\bibfield
  {journal} {\bibinfo  {journal} {PoS}\ }\textbf {\bibinfo {volume}
  {AASKA14}},\ \bibinfo {pages} {037} (\bibinfo {year} {2015})},\ \Eprint
  {http://arxiv.org/abs/1501.00127} {arXiv:1501.00127 [astro-ph.IM]}
  \BibitemShut {NoStop}%
\bibitem [{\citenamefont {Weltman}\ \emph {et~al.}(2020)\citenamefont {Weltman}
  \emph {et~al.}}]{Bull:2018lat}%
  \BibitemOpen
  \bibfield  {author} {\bibinfo {author} {\bibfnamefont {A.}~\bibnamefont
  {Weltman}} \emph {et~al.},\ }\href {\doibase 10.1017/pasa.2019.42} {\bibfield
   {journal} {\bibinfo  {journal} {Publ. Astron. Soc. Austral.}\ }\textbf
  {\bibinfo {volume} {37}},\ \bibinfo {pages} {e002} (\bibinfo {year}
  {2020})},\ \Eprint {http://arxiv.org/abs/1810.02680} {arXiv:1810.02680
  [astro-ph.CO]} \BibitemShut {NoStop}%
\bibitem [{\citenamefont {Seoane}\ \emph {et~al.}(2013)\citenamefont {Seoane}
  \emph {et~al.}}]{eLISA:2013xep}%
  \BibitemOpen
  \bibfield  {author} {\bibinfo {author} {\bibfnamefont {P.~A.}\ \bibnamefont
  {Seoane}} \emph {et~al.} (\bibinfo {collaboration} {eLISA}),\ }\href@noop {}
  {\  (\bibinfo {year} {2013})},\ \Eprint {http://arxiv.org/abs/1305.5720}
  {arXiv:1305.5720 [astro-ph.CO]} \BibitemShut {NoStop}%
\bibitem [{\citenamefont {Amaro-Seoane}\ \emph {et~al.}(2017)\citenamefont
  {Amaro-Seoane} \emph {et~al.}}]{LISA:2017pwj}%
  \BibitemOpen
  \bibfield  {author} {\bibinfo {author} {\bibfnamefont {P.}~\bibnamefont
  {Amaro-Seoane}} \emph {et~al.} (\bibinfo {collaboration} {LISA}),\
  }\href@noop {} {\  (\bibinfo {year} {2017})},\ \Eprint
  {http://arxiv.org/abs/1702.00786} {arXiv:1702.00786 [astro-ph.IM]}
  \BibitemShut {NoStop}%
\bibitem [{\citenamefont {Seto}\ \emph {et~al.}(2001)\citenamefont {Seto},
  \citenamefont {Kawamura},\ and\ \citenamefont {Nakamura}}]{Seto:2001qf}%
  \BibitemOpen
  \bibfield  {author} {\bibinfo {author} {\bibfnamefont {N.}~\bibnamefont
  {Seto}}, \bibinfo {author} {\bibfnamefont {S.}~\bibnamefont {Kawamura}}, \
  and\ \bibinfo {author} {\bibfnamefont {T.}~\bibnamefont {Nakamura}},\ }\href
  {\doibase 10.1103/PhysRevLett.87.221103} {\bibfield  {journal} {\bibinfo
  {journal} {Phys. Rev. Lett.}\ }\textbf {\bibinfo {volume} {87}},\ \bibinfo
  {pages} {221103} (\bibinfo {year} {2001})},\ \Eprint
  {http://arxiv.org/abs/astro-ph/0108011} {arXiv:astro-ph/0108011} \BibitemShut
  {NoStop}%
\bibitem [{\citenamefont {Kawamura}\ \emph {et~al.}(2006)\citenamefont
  {Kawamura} \emph {et~al.}}]{Kawamura:2006up}%
  \BibitemOpen
  \bibfield  {author} {\bibinfo {author} {\bibfnamefont {S.}~\bibnamefont
  {Kawamura}} \emph {et~al.},\ }\href {\doibase 10.1088/0264-9381/23/8/S17}
  {\bibfield  {journal} {\bibinfo  {journal} {Class. Quant. Grav.}\ }\textbf
  {\bibinfo {volume} {23}},\ \bibinfo {pages} {S125} (\bibinfo {year}
  {2006})}\BibitemShut {NoStop}%
\bibitem [{\citenamefont {Hu}\ and\ \citenamefont {Wu}(2017)}]{Hu:2017mde}%
  \BibitemOpen
  \bibfield  {author} {\bibinfo {author} {\bibfnamefont {W.-R.}\ \bibnamefont
  {Hu}}\ and\ \bibinfo {author} {\bibfnamefont {Y.-L.}\ \bibnamefont {Wu}},\
  }\href {\doibase 10.1093/nsr/nwx116} {\bibfield  {journal} {\bibinfo
  {journal} {Natl. Sci. Rev.}\ }\textbf {\bibinfo {volume} {4}},\ \bibinfo
  {pages} {685} (\bibinfo {year} {2017})}\BibitemShut {NoStop}%
\bibitem [{\citenamefont {Luo}\ \emph {et~al.}(2016)\citenamefont {Luo} \emph
  {et~al.}}]{TianQin:2015yph}%
  \BibitemOpen
  \bibfield  {author} {\bibinfo {author} {\bibfnamefont {J.}~\bibnamefont
  {Luo}} \emph {et~al.} (\bibinfo {collaboration} {TianQin}),\ }\href {\doibase
  10.1088/0264-9381/33/3/035010} {\bibfield  {journal} {\bibinfo  {journal}
  {Class. Quant. Grav.}\ }\textbf {\bibinfo {volume} {33}},\ \bibinfo {pages}
  {035010} (\bibinfo {year} {2016})},\ \Eprint
  {http://arxiv.org/abs/1512.02076} {arXiv:1512.02076 [astro-ph.IM]}
  \BibitemShut {NoStop}%
\bibitem [{\citenamefont {Mei}\ \emph {et~al.}(2020)\citenamefont {Mei} \emph
  {et~al.}}]{TianQin:2020hid}%
  \BibitemOpen
  \bibfield  {author} {\bibinfo {author} {\bibfnamefont {J.}~\bibnamefont
  {Mei}} \emph {et~al.} (\bibinfo {collaboration} {TianQin}),\ }\href {\doibase
  10.1093/ptep/ptaa114} {\  (\bibinfo {year} {2020}),\ 10.1093/ptep/ptaa114},\
  \Eprint {http://arxiv.org/abs/2008.10332} {arXiv:2008.10332 [gr-qc]}
  \BibitemShut {NoStop}%
\bibitem [{\citenamefont {Crowder}\ and\ \citenamefont
  {Cornish}(2005)}]{Crowder:2005nr}%
  \BibitemOpen
  \bibfield  {author} {\bibinfo {author} {\bibfnamefont {J.}~\bibnamefont
  {Crowder}}\ and\ \bibinfo {author} {\bibfnamefont {N.~J.}\ \bibnamefont
  {Cornish}},\ }\href {\doibase 10.1103/PhysRevD.72.083005} {\bibfield
  {journal} {\bibinfo  {journal} {Phys. Rev. D}\ }\textbf {\bibinfo {volume}
  {72}},\ \bibinfo {pages} {083005} (\bibinfo {year} {2005})},\ \Eprint
  {http://arxiv.org/abs/gr-qc/0506015} {arXiv:gr-qc/0506015} \BibitemShut
  {NoStop}%
\bibitem [{\citenamefont {Corbin}\ and\ \citenamefont
  {Cornish}(2006)}]{Corbin:2005ny}%
  \BibitemOpen
  \bibfield  {author} {\bibinfo {author} {\bibfnamefont {V.}~\bibnamefont
  {Corbin}}\ and\ \bibinfo {author} {\bibfnamefont {N.~J.}\ \bibnamefont
  {Cornish}},\ }\href {\doibase 10.1088/0264-9381/23/7/014} {\bibfield
  {journal} {\bibinfo  {journal} {Class. Quant. Grav.}\ }\textbf {\bibinfo
  {volume} {23}},\ \bibinfo {pages} {2435} (\bibinfo {year} {2006})},\ \Eprint
  {http://arxiv.org/abs/gr-qc/0512039} {arXiv:gr-qc/0512039} \BibitemShut
  {NoStop}%
\bibitem [{\citenamefont {Harry}\ \emph {et~al.}(2006)\citenamefont {Harry},
  \citenamefont {Fritschel}, \citenamefont {Shaddock}, \citenamefont
  {Folkner},\ and\ \citenamefont {Phinney}}]{Harry:2006fi}%
  \BibitemOpen
  \bibfield  {author} {\bibinfo {author} {\bibfnamefont {G.~M.}\ \bibnamefont
  {Harry}}, \bibinfo {author} {\bibfnamefont {P.}~\bibnamefont {Fritschel}},
  \bibinfo {author} {\bibfnamefont {D.~A.}\ \bibnamefont {Shaddock}}, \bibinfo
  {author} {\bibfnamefont {W.}~\bibnamefont {Folkner}}, \ and\ \bibinfo
  {author} {\bibfnamefont {E.~S.}\ \bibnamefont {Phinney}},\ }\href {\doibase
  10.1088/0264-9381/23/15/008} {\bibfield  {journal} {\bibinfo  {journal}
  {Class. Quant. Grav.}\ }\textbf {\bibinfo {volume} {23}},\ \bibinfo {pages}
  {4887} (\bibinfo {year} {2006})},\ \bibinfo {note} {[Erratum:
  Class.Quant.Grav. 23, 7361 (2006)]}\BibitemShut {NoStop}%
\bibitem [{\citenamefont {Hobbs}\ \emph {et~al.}(2010)\citenamefont {Hobbs}
  \emph {et~al.}}]{Hobbs:2009yy}%
  \BibitemOpen
  \bibfield  {author} {\bibinfo {author} {\bibfnamefont {G.}~\bibnamefont
  {Hobbs}} \emph {et~al.},\ }\href {\doibase 10.1088/0264-9381/27/8/084013}
  {\bibfield  {journal} {\bibinfo  {journal} {Class. Quant. Grav.}\ }\textbf
  {\bibinfo {volume} {27}},\ \bibinfo {pages} {084013} (\bibinfo {year}
  {2010})},\ \Eprint {http://arxiv.org/abs/0911.5206} {arXiv:0911.5206
  [astro-ph.SR]} \BibitemShut {NoStop}%
\bibitem [{\citenamefont {Manchester}(2013)}]{Manchester:2013ndt}%
  \BibitemOpen
  \bibfield  {author} {\bibinfo {author} {\bibfnamefont {R.~N.}\ \bibnamefont
  {Manchester}},\ }\href {\doibase 10.1088/0264-9381/30/22/224010} {\bibfield
  {journal} {\bibinfo  {journal} {Class. Quant. Grav.}\ }\textbf {\bibinfo
  {volume} {30}},\ \bibinfo {pages} {224010} (\bibinfo {year} {2013})},\
  \Eprint {http://arxiv.org/abs/1309.7392} {arXiv:1309.7392 [astro-ph.IM]}
  \BibitemShut {NoStop}%
\bibitem [{\citenamefont {Verbiest}\ \emph {et~al.}(2016)\citenamefont
  {Verbiest} \emph {et~al.}}]{Verbiest:2016vem}%
  \BibitemOpen
  \bibfield  {author} {\bibinfo {author} {\bibfnamefont {J.~P.~W.}\
  \bibnamefont {Verbiest}} \emph {et~al.},\ }\href {\doibase
  10.1093/mnras/stw347} {\bibfield  {journal} {\bibinfo  {journal} {Mon. Not.
  Roy. Astron. Soc.}\ }\textbf {\bibinfo {volume} {458}},\ \bibinfo {pages}
  {1267} (\bibinfo {year} {2016})},\ \Eprint {http://arxiv.org/abs/1602.03640}
  {arXiv:1602.03640 [astro-ph.IM]} \BibitemShut {NoStop}%
\bibitem [{\citenamefont {{Vishniac}}(1982)}]{1982ApJ...257..456V}%
  \BibitemOpen
  \bibfield  {author} {\bibinfo {author} {\bibfnamefont {E.~T.}\ \bibnamefont
  {{Vishniac}}},\ }\href {\doibase 10.1086/160004} {\bibfield  {journal}
  {\bibinfo  {journal} {\apj}\ }\textbf {\bibinfo {volume} {257}},\ \bibinfo
  {pages} {456} (\bibinfo {year} {1982})}\BibitemShut {NoStop}%
\bibitem [{\citenamefont {Rebhan}\ and\ \citenamefont
  {Schwarz}(1994)}]{Rebhan:1994zw}%
  \BibitemOpen
  \bibfield  {author} {\bibinfo {author} {\bibfnamefont {A.~K.}\ \bibnamefont
  {Rebhan}}\ and\ \bibinfo {author} {\bibfnamefont {D.~J.}\ \bibnamefont
  {Schwarz}},\ }\href {\doibase 10.1103/PhysRevD.50.2541} {\bibfield  {journal}
  {\bibinfo  {journal} {Phys. Rev. D}\ }\textbf {\bibinfo {volume} {50}},\
  \bibinfo {pages} {2541} (\bibinfo {year} {1994})},\ \Eprint
  {http://arxiv.org/abs/gr-qc/9403032} {arXiv:gr-qc/9403032} \BibitemShut
  {NoStop}%
\bibitem [{\citenamefont {Weinberg}(2004)}]{Weinberg:2003ur}%
  \BibitemOpen
  \bibfield  {author} {\bibinfo {author} {\bibfnamefont {S.}~\bibnamefont
  {Weinberg}},\ }\href {\doibase 10.1103/PhysRevD.69.023503} {\bibfield
  {journal} {\bibinfo  {journal} {Phys. Rev. D}\ }\textbf {\bibinfo {volume}
  {69}},\ \bibinfo {pages} {023503} (\bibinfo {year} {2004})},\ \Eprint
  {http://arxiv.org/abs/astro-ph/0306304} {arXiv:astro-ph/0306304} \BibitemShut
  {NoStop}%
\bibitem [{\citenamefont {Hook}\ \emph {et~al.}(2021)\citenamefont {Hook},
  \citenamefont {Marques-Tavares},\ and\ \citenamefont {Racco}}]{Hook:2020phx}%
  \BibitemOpen
  \bibfield  {author} {\bibinfo {author} {\bibfnamefont {A.}~\bibnamefont
  {Hook}}, \bibinfo {author} {\bibfnamefont {G.}~\bibnamefont
  {Marques-Tavares}}, \ and\ \bibinfo {author} {\bibfnamefont {D.}~\bibnamefont
  {Racco}},\ }\href {\doibase 10.1007/JHEP02(2021)117} {\bibfield  {journal}
  {\bibinfo  {journal} {JHEP}\ }\textbf {\bibinfo {volume} {02}},\ \bibinfo
  {pages} {117} (\bibinfo {year} {2021})},\ \Eprint
  {http://arxiv.org/abs/2010.03568} {arXiv:2010.03568 [hep-ph]} \BibitemShut
  {NoStop}%
\bibitem [{\citenamefont {Hamaguchi}\ \emph {et~al.}(2014)\citenamefont
  {Hamaguchi}, \citenamefont {Ibe}, \citenamefont {Yanagida},\ and\
  \citenamefont {Yokozaki}}]{Hamaguchi:2014sea}%
  \BibitemOpen
  \bibfield  {author} {\bibinfo {author} {\bibfnamefont {K.}~\bibnamefont
  {Hamaguchi}}, \bibinfo {author} {\bibfnamefont {M.}~\bibnamefont {Ibe}},
  \bibinfo {author} {\bibfnamefont {T.~T.}\ \bibnamefont {Yanagida}}, \ and\
  \bibinfo {author} {\bibfnamefont {N.}~\bibnamefont {Yokozaki}},\ }\href
  {\doibase 10.1103/PhysRevD.90.015027} {\bibfield  {journal} {\bibinfo
  {journal} {Phys. Rev. D}\ }\textbf {\bibinfo {volume} {90}},\ \bibinfo
  {pages} {015027} (\bibinfo {year} {2014})},\ \Eprint
  {http://arxiv.org/abs/1403.1398} {arXiv:1403.1398 [hep-ph]} \BibitemShut
  {NoStop}%
\bibitem [{\citenamefont {Choi}\ \emph {et~al.}(2021)\citenamefont {Choi},
  \citenamefont {Yanagida},\ and\ \citenamefont {Yokozaki}}]{Choi:2020wdq}%
  \BibitemOpen
  \bibfield  {author} {\bibinfo {author} {\bibfnamefont {G.}~\bibnamefont
  {Choi}}, \bibinfo {author} {\bibfnamefont {T.~T.}\ \bibnamefont {Yanagida}},
  \ and\ \bibinfo {author} {\bibfnamefont {N.}~\bibnamefont {Yokozaki}},\
  }\href {\doibase 10.1007/JHEP04(2021)024} {\bibfield  {journal} {\bibinfo
  {journal} {JHEP}\ }\textbf {\bibinfo {volume} {04}},\ \bibinfo {pages} {024}
  (\bibinfo {year} {2021})},\ \Eprint {http://arxiv.org/abs/2012.03266}
  {arXiv:2012.03266 [hep-ph]} \BibitemShut {NoStop}%
\end{thebibliography}%


%

\end{document}